%% file: 0-main-arxiv.tex
\documentclass[12pt]{article}
\usepackage{graphicx} 
\usepackage[table]{xcolor}
\usepackage{pgf}
\usepackage{pgfplotstable}
\usepackage{pgfplots}
\pgfplotsset{compat=1.18}      
\usepackage{collcell}    
\usepackage{array}
\usepackage{booktabs}
\usepackage{float}

\newcommand{\heat}[1]{
  \begingroup
  \pgfmathsetmacro\val{#1}
  \pgfmathsetmacro\pct{min(100,max(0,100*\val))}
  \edef\doCellColor{\noexpand\cellcolor{black!\pct}}
  \doCellColor
  \pgfmathparse{\val>0.5}\ifnum\pgfmathresult=1 \color{white}\fi
  \pgfmathprintnumber[fixed,zerofill,precision=2]{\val}
  \endgroup
}

\newcolumntype{H}{>{\collectcell\heat}r<{\endcollectcell}}
\usepackage{style}
\usepackage{authblk}
\usepackage{amsmath, amssymb, amsthm, bm}
\usepackage{algorithm}
\usepackage{algpseudocode}
\usepackage{caption}  
\usepackage{subcaption}  
\usepackage [english]{babel}
\usepackage [autostyle, english = american]{csquotes}
\usepackage{natbib}
\usepackage{hyperref}
\MakeOuterQuote{"}
\renewcommand{\E}{\mathbb{E}}          
\renewcommand{\Var}{\operatorname{Var}}  
\providecommand{\MSE}{\operatorname{MSE}}

\title{Automated Gating for Flow Cytometry Data Using a Kernel-Smoothed EM Algorithm}

\author[1]{Farhad de Sousa}
\author[2]{François Ribalet}
\author[3]{Jacob Bien}

\affil[1]{Department of Mathematics, University of Southern California}
\affil[2]{School of Oceanography, University of Washington}
\affil[3]{Department of Data Sciences and Operations, University of Southern California}

\begin{document}

\maketitle

\input{01-abstract}
\input{1-Introduction}
\input{2-Method}
\input{3-Theory}
\input{4-Simulations}
\input{5-RealData}
\input{6-Discussions}

\section*{Acknowledgments}
\input{001-acknowledgment}

\bibliographystyle{agsm}
\bibliography{refs}

\newpage 
\input{7-Appendix}

\end{document}

%% file: 01-abstract.tex
\begin{abstract}
Phytoplankton are microscopic algae responsible for roughly half of the world's photosynthesis that play a critical role in global carbon cycles and oxygen production, and measuring the abundance of their subtypes across a wide range of spatiotemporal scales is of great relevance to oceanography. High-frequency flow cytometry is a powerful technique in which oceanographers at sea can rapidly record the optical properties of tens of thousands of individual phytoplankton cells every few minutes. Identifying distinct subpopulations within these vast datasets---a process known as "gating"---remains a major challenge and has largely been performed manually so far. In this paper, we introduce a fast, automated gating method, which accurately identifies phytoplankton populations by fitting a time-evolving mixture of Gaussians model using an expectation-maximization-like algorithm with kernel smoothing. We use simulated data to demonstrate the validity and robustness of this approach, and use oceanographic cruise data to highlight the method's ability to not only replicate but surpass expert manual gating. Finally, we provide the \texttt{flowkernel} R package, written in literate programming, that implements the algorithm efficiently.

\vspace{0.5cm}

{\it Keywords:}  Kernel-smoothing, expectation-maximization, mixture of Gaussians, clustering, phytoplankton
\end{abstract}

%% file: 1-Introduction.tex
\section{Introduction}
\label{sec:Introduction and Motivation}

Marine phytoplankton play a crucial role in Earth's ecosystem, generating nearly half of the planet's oxygen production and organic carbon \citep{Field237}. These microscopic algae form the base of the marine food web and drive global biogeochemical cycles, making it essential to understand how their populations will respond to a changing climate, particularly ocean warming. Measuring the distribution of diverse phytoplankton groups across space and time provides critical information for ultimately understanding their interactions with the environment and how they may  adapt to changing conditions. 

Flow cytometry has changed how oceanographers study phytoplankton by enabling rapid analysis of individual cells. This technique measures the light scattering and fluorescence emitted by cells as they pass through a laser beam, providing information on cell size, internal structure, and pigment composition, such as chlorophyll and phycoerythrin. This data allows researchers to identify and quantify different phytoplankton groups, informing  studies of biogeography, population dynamics, and responses to environmental change \citep{Vaulot1999, Sosik2003, Ribalet2015}. Today, flow cytometers are routinely deployed on oceanographic cruises and at long-term monitoring stations, generating comprehensive datasets that span spatial scales from meters to entire ocean basins. For example, SeaFlow, an automated flow cytometer that continuously samples surface seawater \citep{swalwell2011seaflow}, has produced high-resolution maps of phytoplankton communities in the Northeast Pacific Ocean, providing critical data on their abundance, size, and carbon content \citep{ribalet2019seaflow}.  

Extracting meaningful information from these vast datasets requires identifying and isolating specific phytoplankton populations---a process known as "gating." Traditionally, gating has relied on manual methods, with experts drawing boundaries around populations based on visual inspection of scatter plots and prior knowledge \citep{gating, seaflow-clustering}. This approach is not only time-consuming and laborious but also prone to subjectivity and inconsistencies \citep{flowcore}, especially when analyzing data from different environments where phytoplankton communities and their optical properties vary.  Furthermore, manual gating becomes increasingly impractical with the growing size and complexity of modern flow cytometry datasets due to high-frequency flow cytometry, hindering efficient and comprehensive analysis of phytoplankton dynamics. 

\begin{figure}[t]
  \centering
  \begin{subfigure}[t]{0.26\textwidth}
    \centering
    \includegraphics[width=\linewidth]{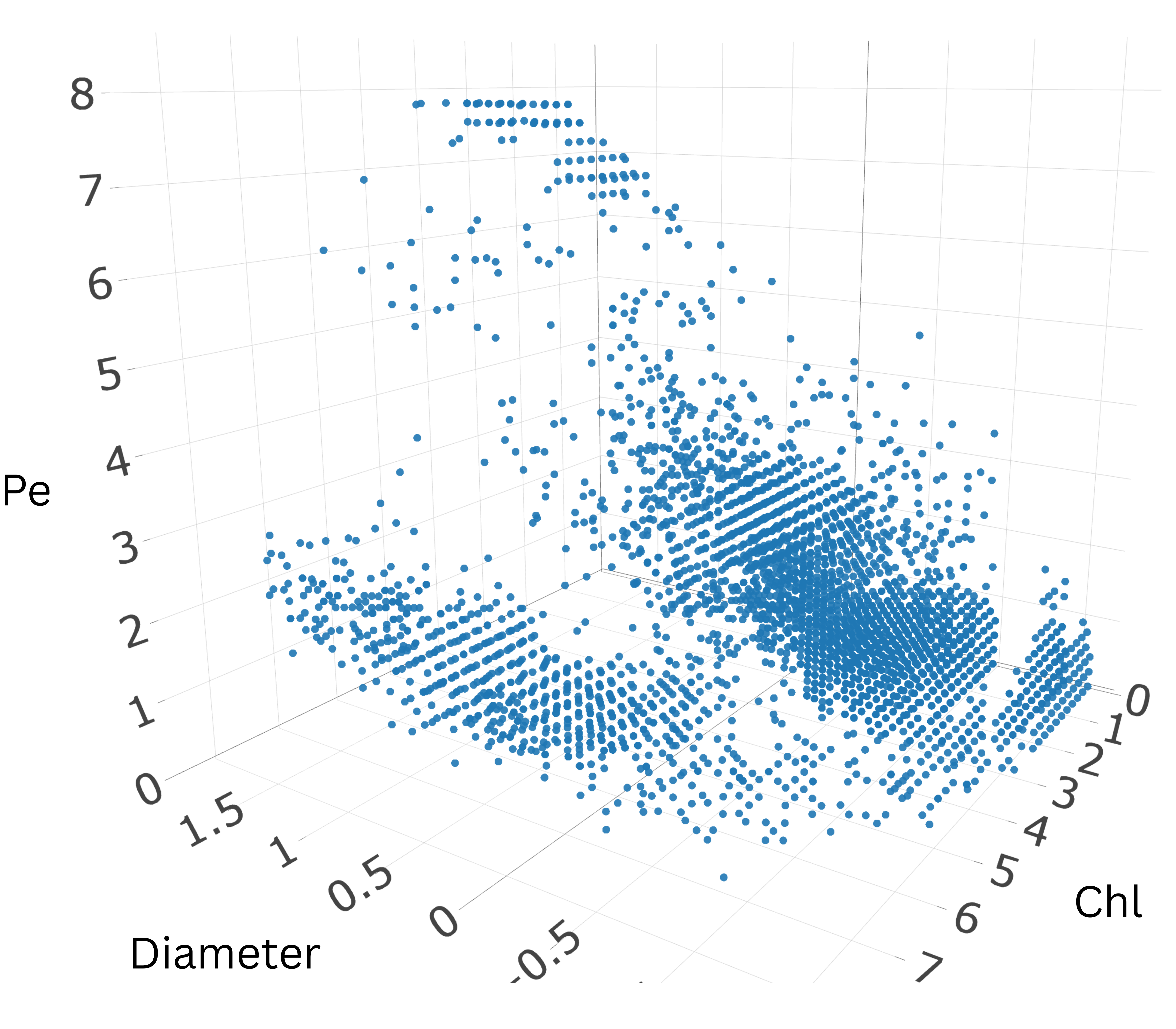}
  \end{subfigure}\hfill
  \begin{subfigure}[t]{0.26\textwidth}
    \centering
    \includegraphics[width=\linewidth]{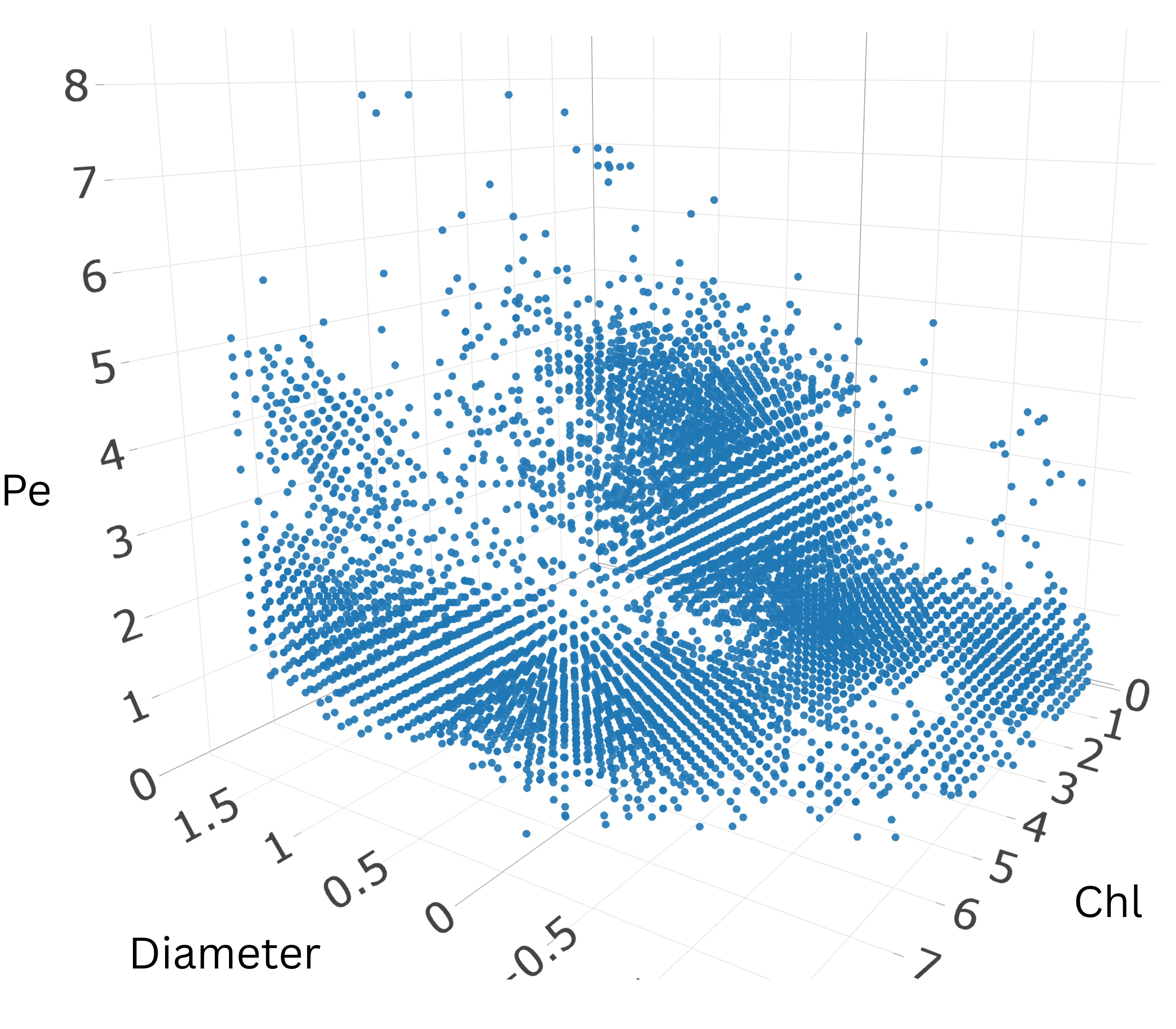}
  \end{subfigure}\hfill
  \begin{subfigure}[t]{0.26\textwidth}
    \centering
    \includegraphics[width=\linewidth]{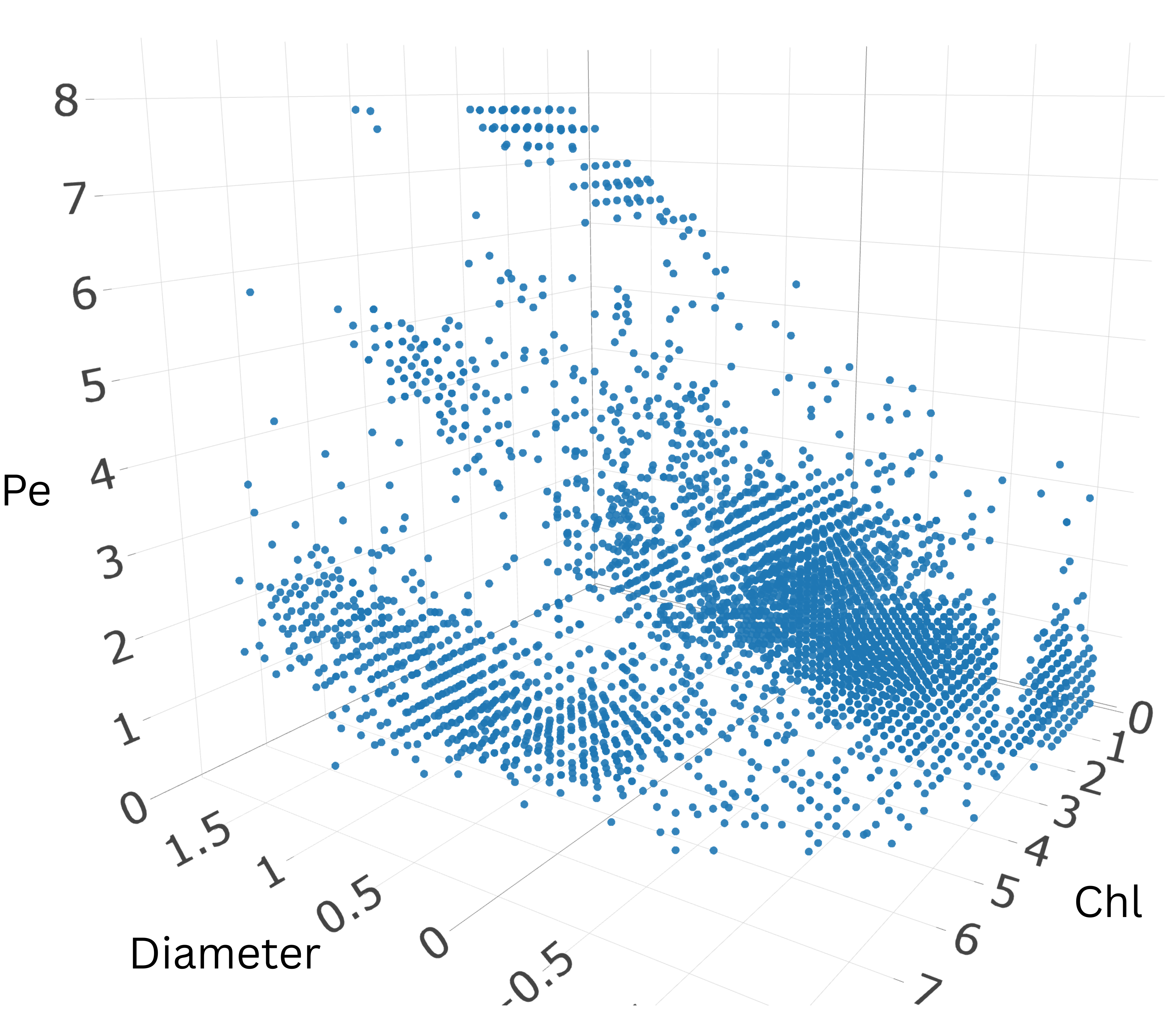}
  \end{subfigure}

  \vspace{0.2em} 

  \begin{subfigure}[t]{0.26\textwidth}
    \centering
    \includegraphics[width=\linewidth]{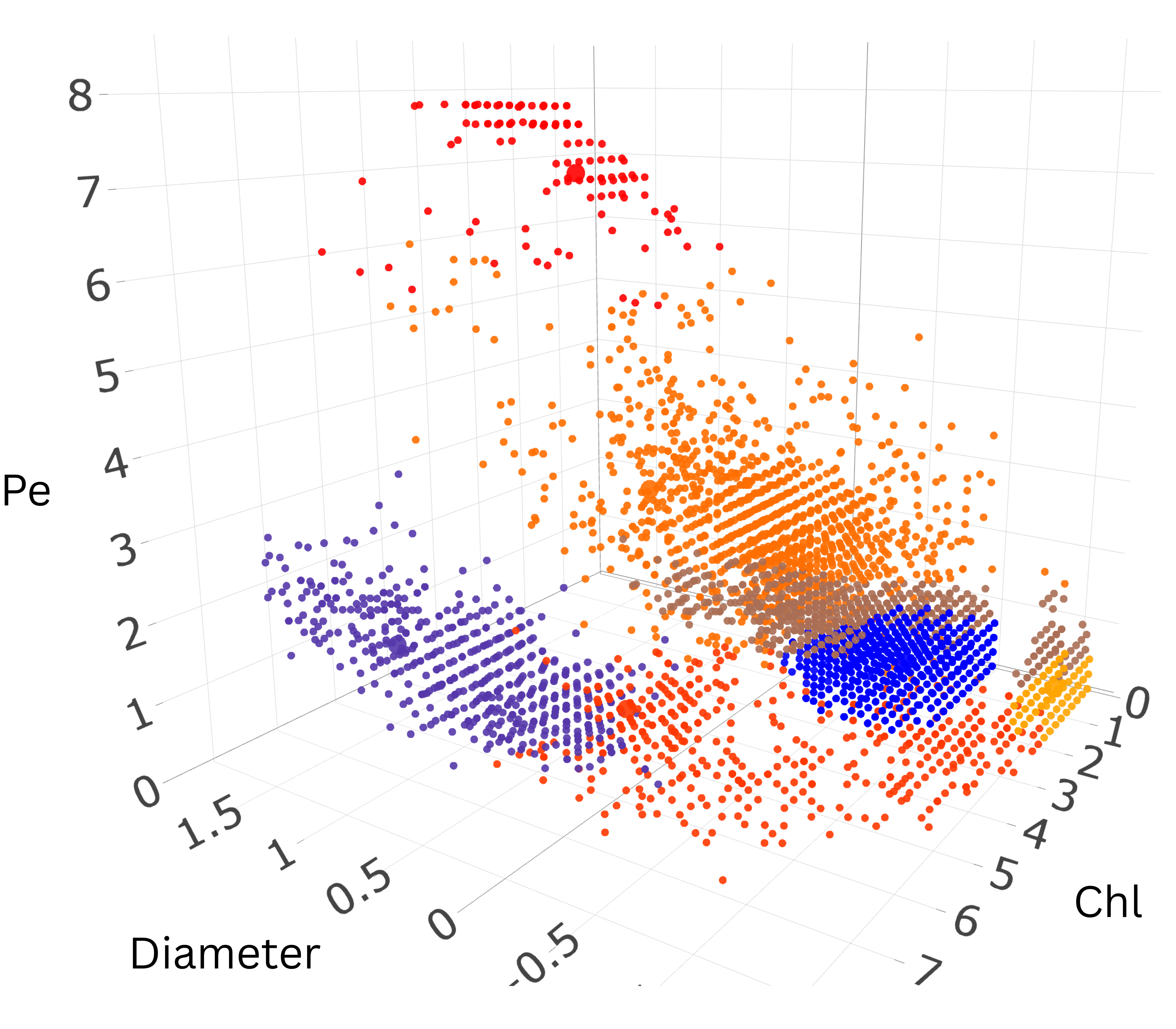}
    \subcaption*{28 May, 21:00 UTC}
  \end{subfigure}\hfill
  \begin{subfigure}[t]{0.26\textwidth}
    \centering
    \includegraphics[width=\linewidth]{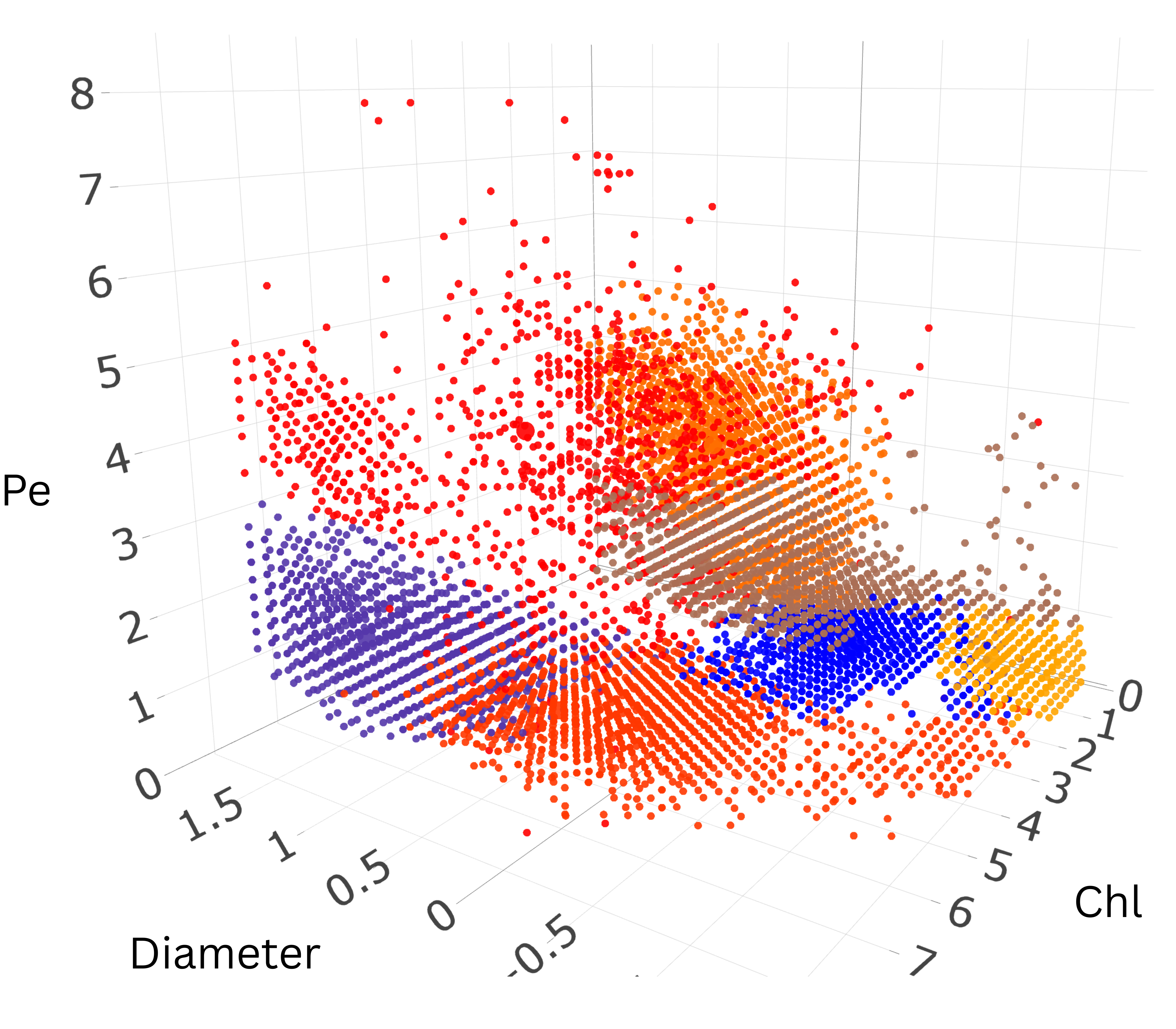}
    \subcaption*{02 June, 01:00 UTC}
  \end{subfigure}\hfill
  \begin{subfigure}[t]{0.26\textwidth}
    \centering
    \includegraphics[width=\linewidth]{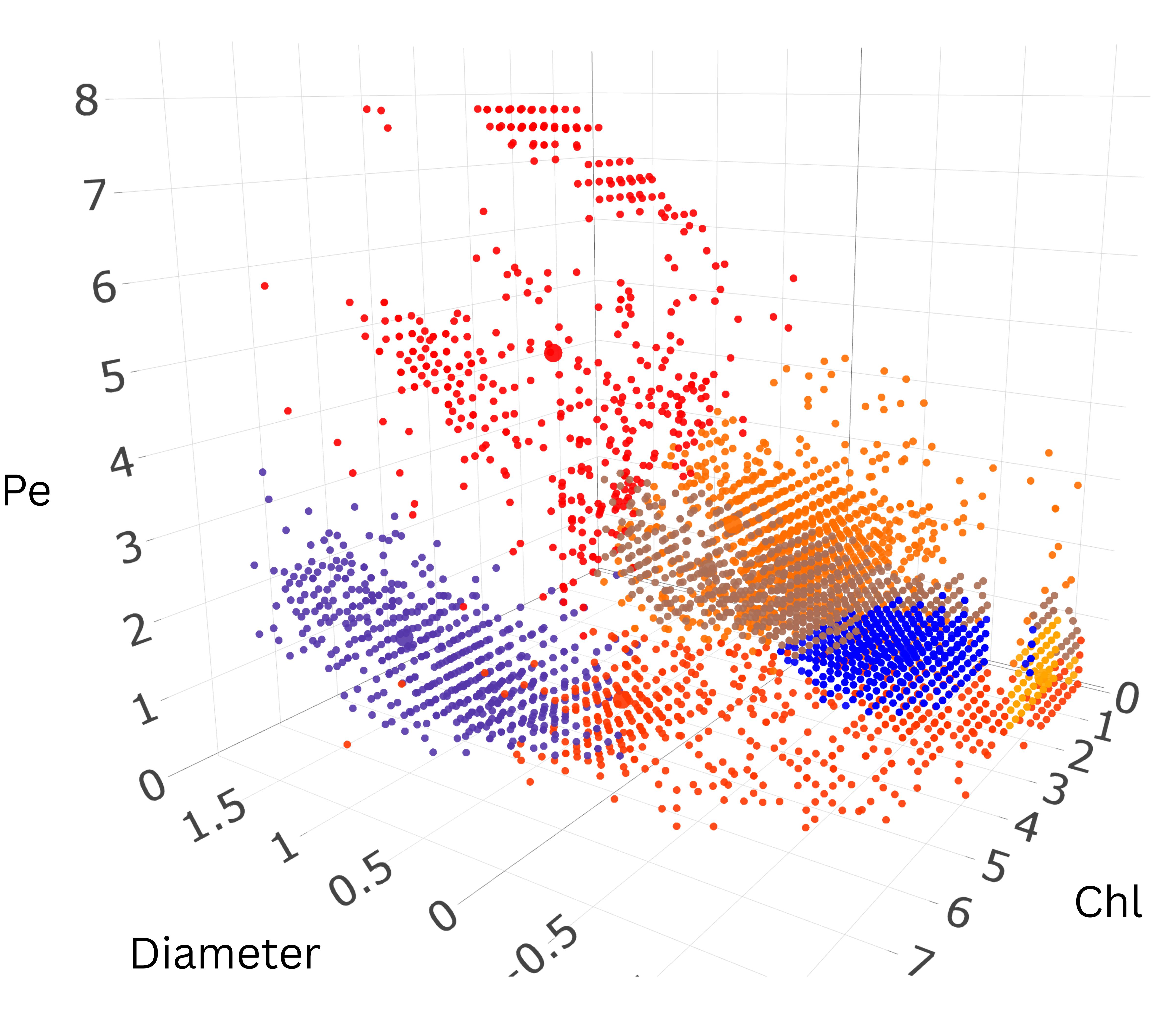}
    \subcaption*{13 June, 07:00 UTC}
  \end{subfigure}

  \caption{Raw and gated (binned) cytograms from a 2017 \textit{SeaFlow} cruise in the Northeast Pacific Ocean. The three dimensions---cell diameter (calculated from light scatter), Chlorophyll content and   Phycoerythrin content---are all on a log scale.}
  \label{fig:raw_gated_cytograms}
\end{figure}

To address the limitations of manual gating, several statistical methods have been employed by the community in recent years \citep{Cheung2021}. The flowMeans \citep{aghaeepour2011rapid}  and flowPeaks \citep{ge2012flowpeaks} algorithms, for example, are part of a broad class of methods that perform gating on a single cytogram at a time.  However, such approaches are not able to leverage the high-frequency setting in which we observe many related cytograms over a period of time. Meanwhile, the CYBERTRACK  \citep{minoura2019cybertrack} algorithm uses information at time $t-1$ to gate at time $t$, but has fixed cluster means and covariances, allowing only cluster proportions to change over time. The flowmix \citep{hyun2023modeling} algorithm allows cluster centers and proportions to change over time too, but requires environmental covariates (like sunlight, sea surface temperature, and salinity) to explain all variability; however, such auxiliary information may not always be available or practitioners may simply not want to assume at the gating stage that environmental covariates are sufficient for explaining cytogram variability. The flowtrend \citep{hyun2025trendfilter} algorithm is a recent algorithm that is closest to our proposal. This method, like ours, uses cytometry data from all time points for gating at each time point and allows for changing means and proportions. While flowtrend is effective at gating, it is computationally expensive because it requires solving a sequence of convex optimization problems.  This in turn limits the ease with which the method can be applied to large-scale high-frequency cytometry datasets considered in this paper. 

Our approach, which we call \textbf{flowkernel}, is designed to be a fast and light-weight solution.  Like flowtrend, it uses data from multiple time points to gate each cytogram, but it replaces convex optimization problems with much simpler updates, making it a practical alternative for oceanographers. Flowkernel fits a time-evolving mixture of Gaussians model to flow cytometry data using a kernel-smoothed expectation-maximization-like (EM) algorithm. The method is effective at capturing continuous shifts in phytoplankton populations observed in real-world data, offering a robust and accurate representation of population dynamics over time. In effect, it provides a more objective and reproducible approach to identifying phytoplankton populations than manual gating, while still being fast enough to apply to large modern datasets.

We provide an implementation of our algorithm in the \texttt{flowkernel} R package, which was written in literate programming \citep{Knuth1992, litr}, showing all the source code of the package along with full explanations of the method in a single bookdown document. 

%% file: 2-Method.tex
\section{Methodology}
\label{sec:method}

High-frequency flow cytometry data can be represented as a time series of scatterplots, $ Y_1, \ldots, Y_T $, where each $ Y_t $ is an $ n_t \times d $ matrix, referred to as a cytogram. Each row in $ Y_t $ represents a vector $ Y_{it} \in \mathbb{R}^d $, for $ i = 1, \ldots, n_t $, that captures the optical properties of the $i$th observed particle at time $ t $. Each dimension in the $d$-dimensional scatterplot $Y_t$ represents an optical property that aids in differentiating between various cell types.

In continuous-time flow cytometry data, we notice that each scatterplot looks approximately like a mixture of Gaussians, and the general clustering structure seen in each scatterplot is slowly varying over time. To model data like this, we wish to fit a smoothly-varying mixture of Gaussians model, where $(\mu_{kt},\Sigma_{kt},\pi_{kt})$ are parameters that are slowly varying in $t$:

\begin{equation}\label{eq:mixture-model}
    Y_{it}|\{Z_{it}=k\}\sim N_d(\mu_{kt},\Sigma_{kt}),\qquad\mathbb P(Z_{it}=k)=\pi_{kt}.
\end{equation}

Our method, sketched in Algorithm~\ref{alg:kernel-em} and presented in greater detail below, follows the precedent of modifying the EM algorithm to address specific modeling challenges, as seen in prior work such as \citet{green1990penalize}. While the standard EM algorithm focuses solely on maximizing the likelihood of the observed data, we incorporate kernel smoothing to tie different time points together and ensure that parameters evolve smoothly over time.

\begin{algorithm}[t]
\caption{Kernel-Smoothed EM Algorithm Overview}\label{alg:kernel-em}
\begin{algorithmic}
\State \textbf{Initialize} parameters $\mu_{tk}, \pi_{tk}, \Sigma_{tk}$, $\forall t$ (time points) and $\forall k$ (clusters)
\For{iteration in \textit{iterations}}
    \State \textbf{E-step}: Calculate responsibilities $\hat\gamma_{itk}=\hat{\mathbb{P}}(Z_{it}=k|Y_{it})$, using Equation~\eqref{eq:E-step}
    \State \textbf{Kernel-Smoothed M-step}: Update parameters $\hat{\mu}_{tk}, \hat{\pi}_{tk}, \hat{\Sigma}_{tk}$ using the kernel-smoothed maximum likelihood estimates in Equations~\eqref{eq:M-step-pi}, ~\eqref{eq:M-step-mu}, and ~\eqref{eq:M-step-Sigma}
\EndFor
\end{algorithmic}
\end{algorithm}

We primarily work with \textit{binned} flow cytometry data, where the three-dimensional space is divided into a grid, and $Y_{it}$ represents the location of the $i$th bin at time $t$. For each bin at each point in time, we have an associated weight, $C _i^{(t)}$. Rather than using simple particle counts as multiplicities, we employ carbon biomass as the weighting factor for each bin. Phytoplankton populations in the ocean exhibit highly imbalanced particle distributions, with small abundant populations vastly outnumbering larger, less abundant ones. Biomass weighting helps balance this distribution by giving appropriate weight to larger cells.

We begin by describing the two main components of the algorithm---the E-step and the M-step---before presenting our approach to initialization.

\subsection{E-step}
Given provisional estimates of $(\mu_{tk}, \pi_{tk}, \Sigma_{tk})$ for all time points $t$ and all clusters $k$, the E-step computes for each $Y_{it}$ how "responsible" each cluster is for it. In particular, the responsibility vector $(\hat\gamma_{it1},\ldots,\hat\gamma_{itK})$ is a probability vector, and so we are giving each point a "soft" assignment, as opposed to a "hard" assignment where each point is completely assigned to a single cluster.  The responsibilities are computed using Bayes rule:

\begin{equation}\label{eq:E-step}
\hat\gamma_{itk}=\hat{\mathbb{P}}(Z_{it}=k|Y_{it})=\frac{\hat \pi_{tk}\phi(Y_{it};\hat\mu_{tk},\hat\Sigma_{tk})}{\sum_{\ell=1}^K\hat \pi_{t\ell}\phi(Y_{it};\hat\mu_{t\ell },\hat\Sigma_{t\ell})},
\end{equation}
where $\phi$ denotes the multivariate normal density. 

This E-step matches precisely the standard E-step for fitting mixture of Gaussians (see, e.g., \citealt{hastie2009elements}).

\subsection{Kernel-Smoothed M-step}
In the kernel-smoothed M-step, we update the estimates of our parameters $\mu_{tk},\Sigma_{tk}$, and $\pi_{tk}$. As discussed above, we are working with binned data, where $C_i^{(t)}$ is the biomass at time $t$ in bin $i$, for $i=1,\ldots, B$, where $B$ is the total number of bins. (For unbinned data, we take $C_i^{(t)} = 1$ for all $i$ and $t$.) We define $n_s = \sum_{i=1}^{B} C_i^{(s)}$ to be the total biomass at time $s$ and take
\begin{equation}
\nonumber
\hat n_{sk}=\sum_{i=1}^{B}C_i^{(s)}\hat\gamma_{isk},
\end{equation} 
as an estimate of the total biomass in class $k$ at time $s$. Given a kernel function $w_{h_\pi}(\cdot)$ with bandwidth $h_\pi$, we then estimate $\pi_{tk}$ via

\begin{equation}\label{eq:M-step-pi}
\hat\pi_{tk}=\frac{\sum_{s=1}^{T}w_{h_\pi}(t-s)\hat n_{sk}}{\sum_{s=1}^{T}{w_{h_\pi}(t-s)n_s}}.
\end{equation} \\
This is equivalent to dividing the Nadaraya-Watson kernel-smoothed regression estimate of the total biomass in cluster $k$ at time $t$ by the kernel-smoothed regression estimate of the total biomass at time $t$.

For $\mu_{tk}$, we first compute the biomass-weighted sum of all points "softly" assigned to cluster $k$ at each time $s$:

\begin{equation}
\nonumber
\sum_{i=1}^{B}C_i^{(s)}\hat\gamma_{isk}Y_{is}.
\end{equation}
We then proceed to use this in our smoothed estimate: 

\begin{equation}\label{eq:M-step-mu}
\hat\mu_{tk}=\frac{\sum_{s=1}^{T}w_{h_\mu}(t-s)\sum_{i=1}^{B}C_i^{(s)}\hat\gamma_{isk}Y_{is}}{\sum_{s=1}^{T}w_{h_\mu}(t-s)\hat n_{sk}}.
\end{equation} \\
The denominator above is like that in \eqref{eq:M-step-pi}, except that we use a different bandwidth $h_\mu$ for our smoothed estimated for the total biomass (or number of points) at time $t$.  

Similarly for $\Sigma_{tk}$, we first compute a weighted outer product at each time point, and then smooth over time:

\begin{equation}\label{eq:M-step-Sigma}
\hat\Sigma_{tk}=\frac{\sum_{s=1}^{T}w_{h_\Sigma}(t-s)\sum_{i=1}^{B}C_i^{(s)}\hat\gamma_{isk}(Y_{is}-\hat\mu_{sk})(Y_{is}-\hat\mu_{sk})^\top}{\sum_{s=1}^{T}w_{h_\Sigma}(t-s)\hat n_{sk}}.
\end{equation}

Note that the M-step equations can be applied for any $t$, not just those time points where data are available. This allows us to predict parameter values at times without data, as done in the cross-validation procedure described below.

\subsection{Cross Validation}

Our estimator has three kernel-smoother bandwidths $(h_\pi,h_\mu,h_\Sigma)$. We implement a 5-fold cross-validation scheme similar to \citet{zou2020changepoint}, in which fold $\mathcal I_\ell$ (for $\ell=1,\ldots,5$) constitutes every fifth time point, so $\mathcal{I}_\ell = \{\ell, \ell + 5, \ell + 10, \dots\}$.

We first apply Algorithm~\ref{alg:kernel-em} on all $\{Y_{s}:s\not\in\mathcal I_\ell\}$, which yields parameter estimates
$$\hat\theta_s^{-\mathcal I_\ell}:=\left(\{\hat\gamma^{-\mathcal I_\ell}_{isk}\}_{ik},\{\hat\pi_{sk}^{-\mathcal I_\ell}\}_{k},\{\hat\mu^{-\mathcal I_\ell}_{sk}\}_k,\{\hat\Sigma^{-\mathcal I_\ell}_{sk}\}_k\right)$$
for all $s\not\in \mathcal I_\ell$, where the superscript $-\mathcal I_\ell$ emphasizes that data at times in $\mathcal I_\ell$ are not used.  We further use the shorthand $\hat\theta^{-\mathcal I_\ell}=\{\hat\theta_s^{-\mathcal I_\ell}:s\not\in\mathcal I_\ell\}$.  We next estimate $\pi_{tk}$, $\mu_{tk}$, and $\Sigma_{tk}$ for times $t\in\mathcal{I}_\ell$, by using \eqref{eq:M-step-pi}--\eqref{eq:M-step-Sigma}.  We denote these estimates $\hat\pi_{kt}(\hat\theta^{-\mathcal I_\ell})$, $\hat\mu_{kt}(\hat\theta^{-\mathcal I_\ell})$, and $\hat\Sigma_{kt}(\hat\theta^{-\mathcal I_\ell})$.

Finally, we evaluate the log-likelihood on these unseen time points $\mathcal{I}_\ell$ using the estimated parameters. We repeat this procedure for the remaining folds and take the average to give us a final cross-validated likelihood for the current set of parameter values:

\begin{equation}\label{eq:cv}
\text{CV} = \frac{1}{5}\sum_{\ell=1}^{5} \frac{1}{|\mathcal{I}_\ell|} \sum_{t \in \mathcal{I}_\ell} \sum_{i \in [n_t]} \log\mathcal{L}\left(Y_{it}; \left\{\left(\hat\pi_{tk}(\hat\theta^{-\mathcal I_\ell}),\hat\mu_{tk}(\hat\theta^{-\mathcal I_\ell}), \hat\Sigma_{tk}(\hat\theta^{-\mathcal I_\ell})\right)\right\}_k\right).
\end{equation}

When applying our method to the 2017 SeaFlow cruise described in Section~\ref{sec:realdata}, we chose to vary $h_{\mu}$ and $h_{\pi}$ over a grid of values, while keeping $h_{\Sigma}$ fixed. Our grid values for $h_{\mu}$ and $h_{\pi}$ varied on a logarithmic scale, starting at 1 hour and ending at the the total length of time for the data, which was about 350 hours. With these choices, setting $h_\Sigma = 15$, and running CV over a $7\times7$ grid, cross validation chose $h_\mu = 23$ and $h_\pi = 108$. 

\subsection{Initialization}
\label{subsection:const-init}

In order to carry out the first E-step, we need estimates of $\mu_{tk},\Sigma_{tk}$, and $\pi_{tk}$ for every cluster at every time point. Our primary approach to a quick initialization is to choose $\mu_{tk},\Sigma_{tk}$, and $\pi_{tk}$ that are constant over time. We do this by sampling a subset of time points from the data uniformly at random, and for each selected time point, we sample data points $Y_{it}$ uniformly at random again. Combining these sampled data points into a single cytogram, we fit a single mixture of Gaussians using a standard EM algorithm and use the resulting parameters as time-constant initial values for the kernel-EM algorithm. While implementing the algorithm on the 2017 SeaFlow cruise, we randomly sampled 50 time points, and then sampled 50 data points at each of these time points, and finally used the standard EM algorithm to fit a mixture of Gaussians to the combined 2500 points.

We also developed a Bayesian initialization scheme that allows the parameters to change over time, but the constant initialization is more computationally efficient, and produces results---at the end of the kernel-EM algorithm---comparable to the Bayesian initialization, and so we choose the constant method in practice. The Bayesian initialization scheme is described in the appendix in Section~\ref{subsec:bayesian_init}.

%% file: 3-Theory.tex
\section{Theory}
\label{sec:Theory}

\begin{figure}[t]
    \centering
    \begin{minipage}{0.32\textwidth}
        \centering
        \includegraphics[width=\textwidth]{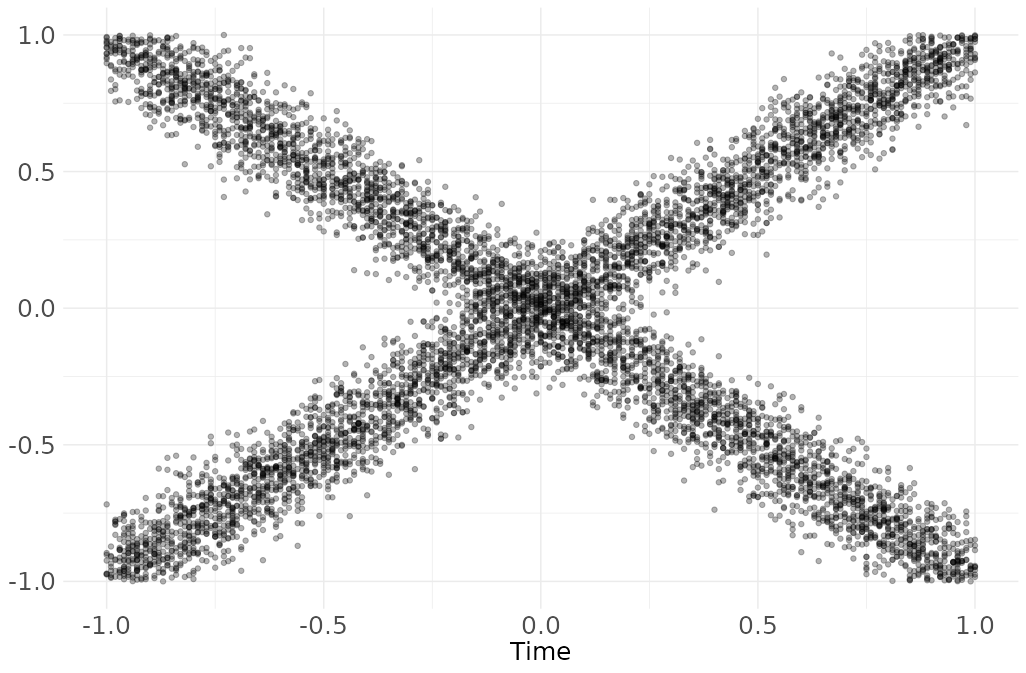}
        \subcaption{Raw Data}
        \label{fig:raw-data}
    \end{minipage}
    \hfill
    \begin{minipage}{0.32\textwidth}
        \centering
        \includegraphics[width=\textwidth]{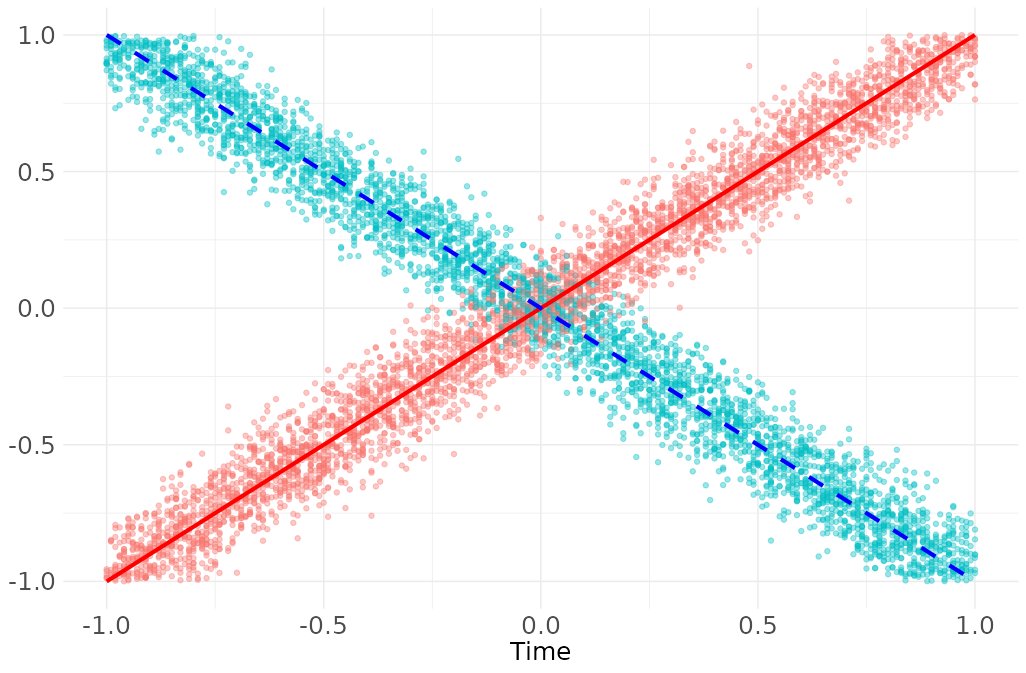}
        \subcaption{$z^{lin}$ and $\mu^{lin}$}
        \label{fig:linmeans}
    \end{minipage}
    \hfill
    \begin{minipage}{0.32\textwidth}
        \centering
        \includegraphics[width=\textwidth]{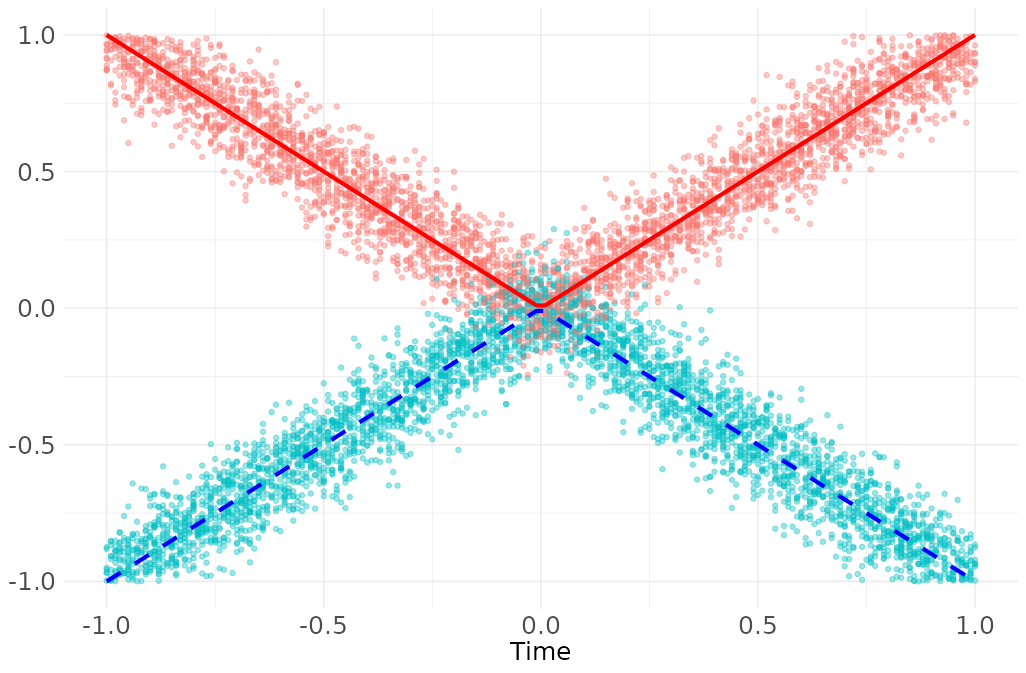}
        \subcaption{$z^{AV}$ and $\mu^{AV}$}
        \label{fig:absmeans}
    \end{minipage}
    \caption{Two alternative cluster assignments for this non-identifiable cluster crossing data.}
    \label{fig:crossing_comparison}
\end{figure}

In this section, we investigate the interplay between smoothness and mixture modeling in our estimator, as it relates to cross validation.  In particular, whenever two or more clusters cross each other, there is inherent non-identifiability that arises.  To study what happens in such a setting, we consider a simple example, shown in Figure~\ref{fig:crossing_comparison}.  Figure~\ref{fig:raw-data} depicts data that could equally well have arisen from either of the two mixture models in Figures~\ref{fig:linmeans}~or~\ref{fig:absmeans}. The two sets of cluster mean functions $(\mu^{lin}_1(t),\mu^{lin}_2(t))$ and $(\mu^{AV}_1(t),\mu^{AV}_2(t))$, while giving rise to identical data-generating mechanisms, differ in smoothness---the first being two linear functions whereas the second are both non-smooth at 0. Our intuition, which we formalize in this section, is that cross validation of our method will tend to favor smoother solutions.  In particular, we will show that the smoother description results in more accurate estimation and prediction.

At each time point $ t \in [-1, 1]$, we have observations $ Y_{it} $ from two clusters 
generated according to the model
\begin{equation}
Y_{it} \mid \{ Z_{it} = k \} \sim \mathcal{N}\big( \mu_k(t),\, \sigma^2 \big), \quad \text{for } k \in \{0, 1\},
\end{equation}
where $ Z_{it} $ denotes the true cluster assignment of observation $ i $ at time $ t $, and $ \mu_k(t) $ is the true mean function of cluster $ k $ at time $ t $.  We assume each cluster has exactly $n$ points at each time point. As shown in the coloring of points in Figures~\ref{fig:linmeans}~or~\ref{fig:absmeans}, the two different mean functions have correspondingly different $Z_{it}$'s, which we denote by $z^{lin}$ and $z^{AV}$. 

We define the \emph{oracle} version of our estimator to be $\hat{\mu}_k(t; Y, z)$, where $z$ are the true cluster assignments.  In words, the oracle uses our algorithm's M-step, but following an idealized E-step in which (soft) responsibilities are replaced by true (hard) cluster memberships:
\begin{equation}
\hat{\mu}_k(t; Y, z) = \frac{\displaystyle \sum_{s=1}^T w_{h}(t - s) \sum_{i=1}^{n_s} \mathbf{1}\{ z_{is} = k \} Y_{is}}{\displaystyle \sum_{s=1}^T w_{h}(t - s) \sum_{i=1}^{n_s} \mathbf{1}\{ z_{is} = k \}}, \quad \text{for } k = 0, 1.
\end{equation}
Our goal is to compare the performance of the oracle estimators $ \hat{\mu}_k(t; Y, z^{lin}) $ and $ \hat{\mu}_k(t; Y, z^{AV}) $, corresponding to the mean functions $ \mu_k^{lin}(t) $ and $ \mu_k^{AV}(t) $, respectively.

\begin{theorem}
\label{thm:mse_comparison}
Assuming the number of observations in the two clusters at each time point are equal and constant across all $t$, the oracle estimator associated with the linear means $ \mu_k^{lin}(t) $ has a lower mean squared error than the one associated with the absolute value means $ \mu_k^{AV}(t) $. Specifically, for all $ t \in [-1, 1] $,
\begin{equation}
\label{eq:mse_inequality}
\mathbb{E}\left[ \sum_{k=0}^{1} \left( \hat{\mu}_k(t; Y, z^{lin}) - \mu_k^{lin}(t) \right)^2 \right] \leq \mathbb{E}\left[ \sum_{k=0}^{1} \left( \hat{\mu}_k(t; Y, z^{AV}) - \mu_k^{AV}(t) \right)^2 \right].
\end{equation}
Moreover, if $ \tilde{Y}_t $ is a new test point at time $ t $, with $ \mathbb{P}(\tilde{Z}_t = 0) = \mathbb{P}(\tilde{Z}_t = 1) = \tfrac{1}{2} $, then the oracle estimator $ \hat{\mu}_k(t; Y, z^{lin}) $ also has a lower expected prediction error:
\begin{equation}\label{eq:epe-lin-vs-av}
\mathbb{E}\left[ \sum_k\left( \tilde{Y}_t - \hat{\mu}_{k}(t; Y, z^{lin}) \right)^2 \right] \leq \mathbb{E}\left[ \sum_k\left( \tilde{Y}_t - \hat{\mu}_{k}(t; Y, z^{AV}) \right)^2 \right].
\end{equation}
\end{theorem}
\vspace{2mm}

Equation~\eqref{eq:epe-lin-vs-av} tells us that the estimator based on the linear cluster assignments predicts better on new data than the estimator based on the absolute value cluster assignments.  This suggests that if one were to use cross validation to decide between these two models, the linear one would be favored.

The full proof of Theorem~\ref{thm:mse_comparison} is given in Section~\ref{subsec:thm1-proof} of the appendix; we present only a sketch here. The key idea 
of the proof is that, in the linear scenario, the oracle estimator is unbiased because kernel smoothing places equal weight on observations around the point being estimated. This symmetry—under the assumption that each cluster has the same number of points—keeps the estimator centered on the true cluster means. The mean squared error (MSE) is therefore just equal to the variance of the estimator.

In the absolute value scenario, however, the absolute value function induces bias via its convexity or concavity. By Jensen’s inequality, kernel smoothing cannot perfectly follow abrupt changes, causing the estimator to “fill the valley” or “trim the hill.” This bias shifts the estimator away from the true cluster means, thereby increasing the MSE compared to the linear scenario.

%% file: 4-Simulations.tex
\section{Simulations}

To assess the effectiveness of the kernel-EM algorithm, we compare it in simulation to two natural baseline methods.  The first baseline method takes data from all time points and collapses them into one large aggregated cytogram and then fits a single Gaussian mixture model on this cumulative data to give a constant fit over time.  The second baseline method fits a separate Gaussian mixture model at each time point, and the different cluster labels between time points are matched together using the Hungarian algorithm \citep{kuhn1955hungarian}. 

We explore two scenarios in our simulations meant to push our method to challenging extremes: one in which a cluster disappears for a while and then reappears, and one in which clusters have significant overlap with each other. Since the true cluster labels are available in these simulation scenarios, we can evaluate the quality of our fit using the Rand index \citep{rand-index}, which quantifies the similarity between two clusterings of the same dataset. To assess clustering performance across all time points, we compute the Rand index at each time point, using the true cluster labels and cluster labels that are sampled based on the fitted responsibilities, and then take an average over all time points.

\begin{figure}[t]
    \centering
    \begin{subfigure}[b]{0.32\textwidth}
        \centering
        \includegraphics[width=\textwidth]{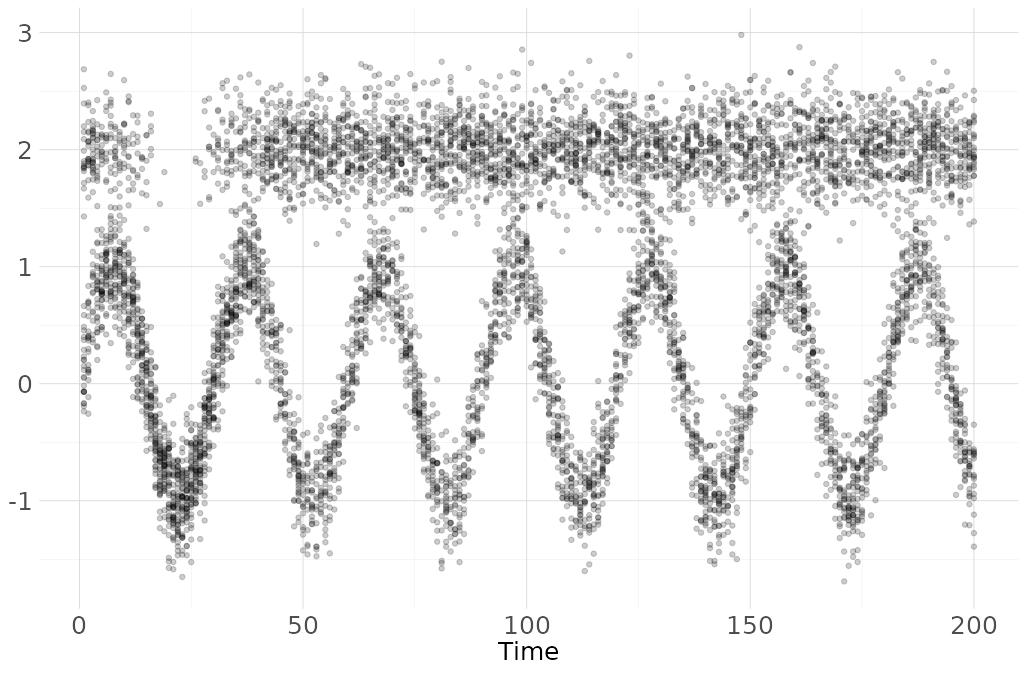}
        \caption{Duration = 5}
    \end{subfigure}
    \begin{subfigure}[b]{0.32\textwidth}
        \centering
        \includegraphics[width=\textwidth]{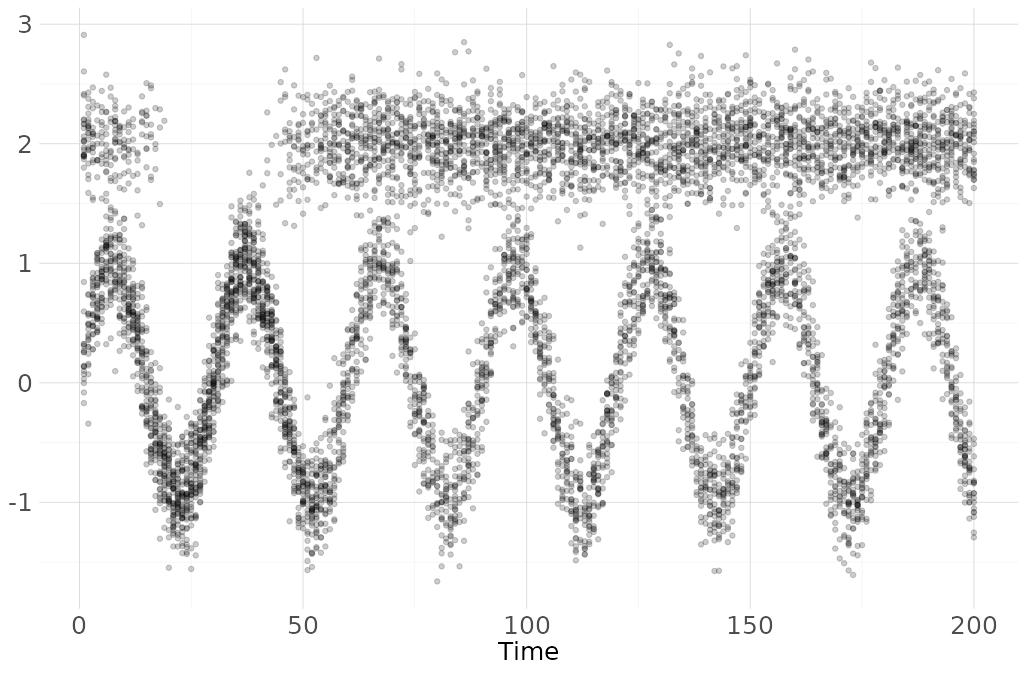}
        \caption{Duration = 20}
    \end{subfigure}
    \begin{subfigure}[b]{0.32\textwidth}
        \centering
        \includegraphics[width=\textwidth]{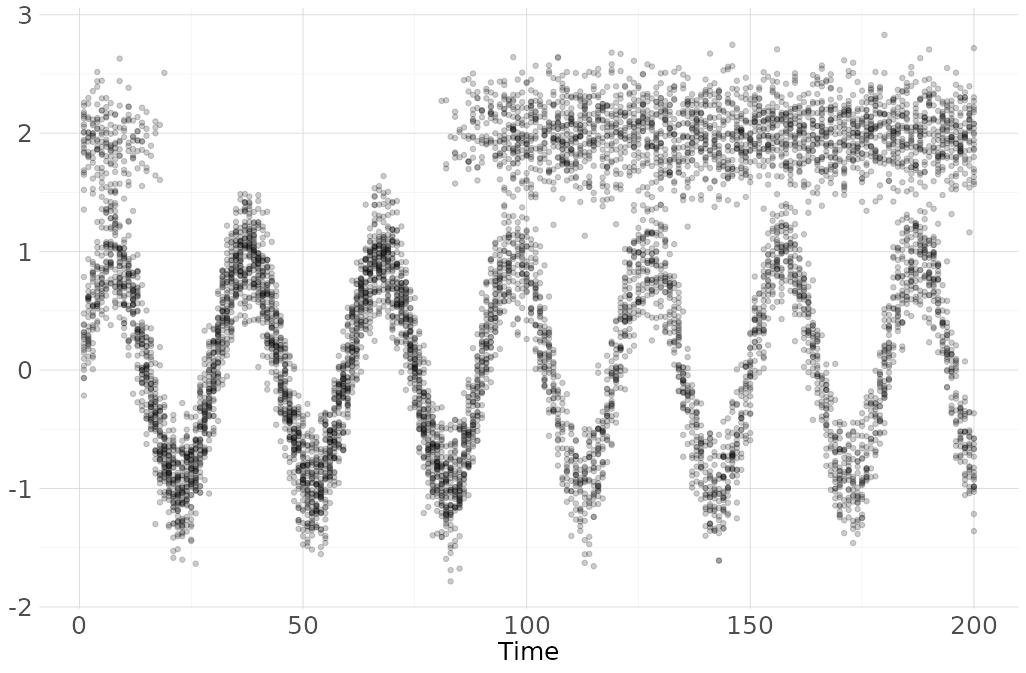}
        \caption{Duration = 60}
    \end{subfigure}

    \caption{We simulate a gradually increasing disappearance duration in the 1-d case.}
    \label{fig:dis-duration-raw}
\end{figure}

\begin{figure}[t] 
    \centering
    \begin{subfigure}[t]{0.32\textwidth}
        \centering
        \includegraphics[width=\textwidth]{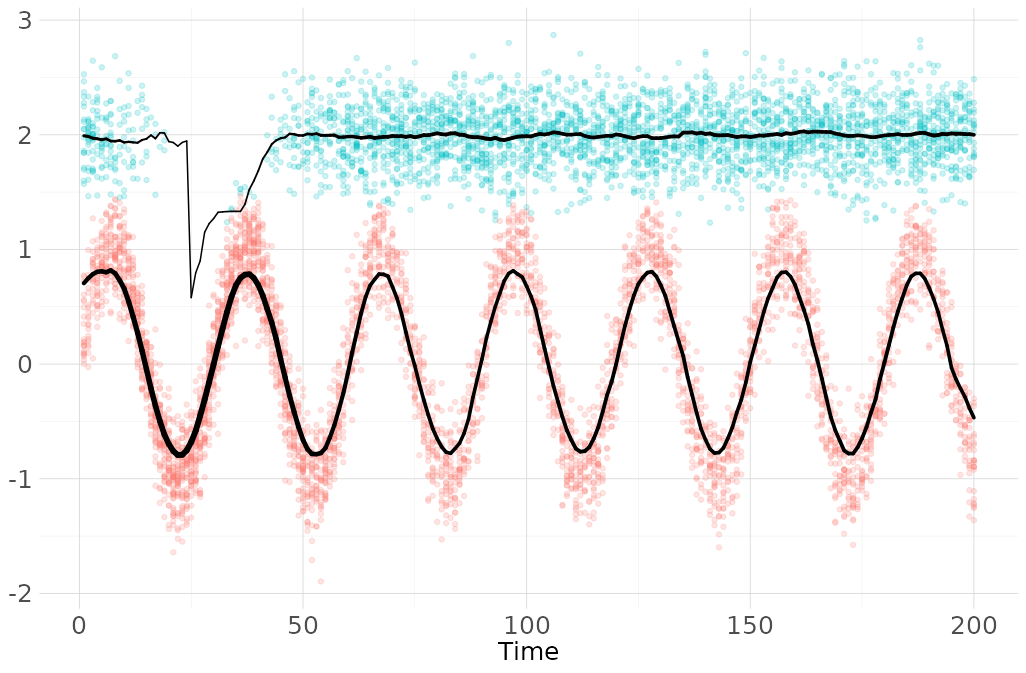}
        \caption{Kernel-EM fit}
        \label{fig:kernel_fit_dis_20}
    \end{subfigure}
    \hfill
    \begin{subfigure}[t]{0.32\textwidth}
        \centering
        \includegraphics[width=\textwidth]{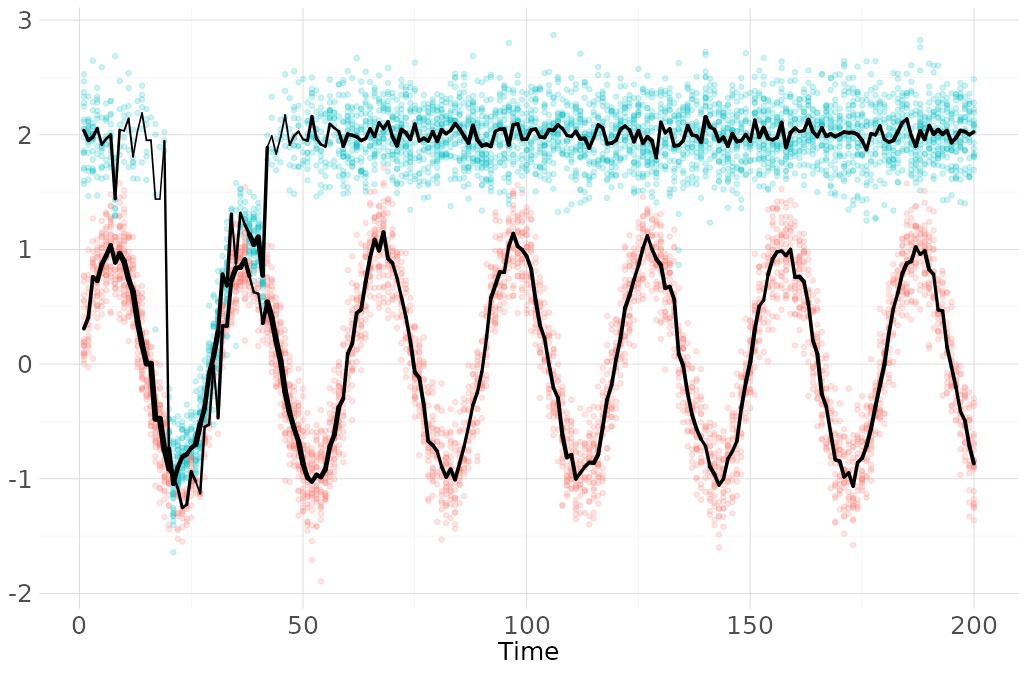}
        \caption{Hungarian fit}
        \label{fig:hung_fit_dis_20}
    \end{subfigure}
    \hfill
    \begin{subfigure}[t]{0.32\textwidth}
        \centering
        \includegraphics[width=\textwidth]{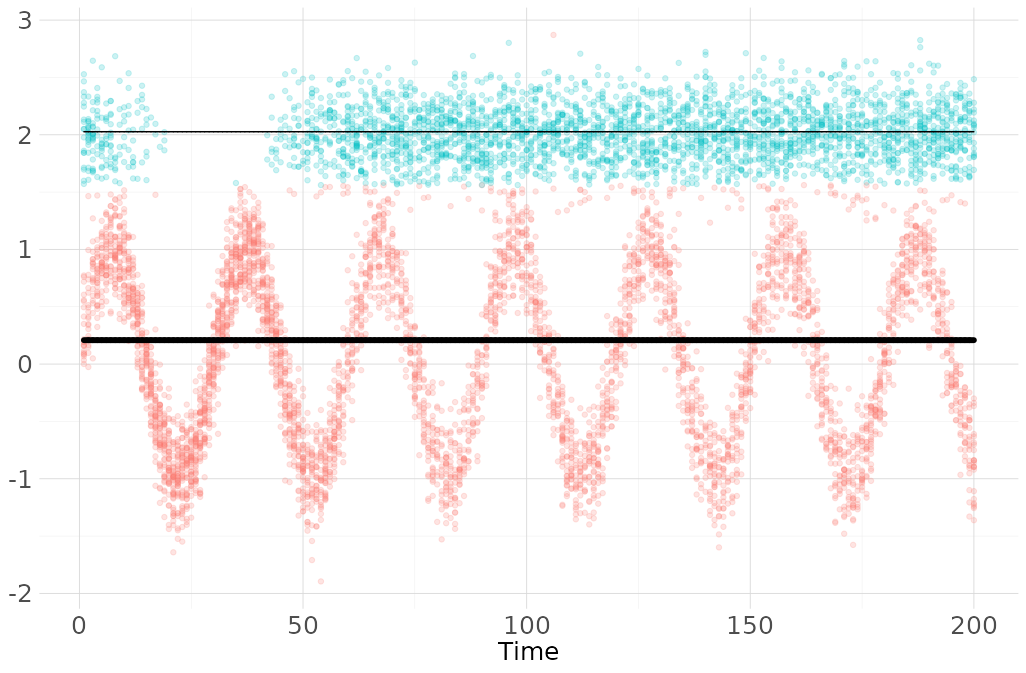}
        \caption{Constant fit}
        \label{fig:const_fit_dis_20}
    \end{subfigure}

    \caption{Comparison of the three different methods at disappearance duration 20. The thickness of the cluster mean lines indicates $\hat{\pi}_{tk}$ at each time point.} 

    \label{fig:disappearance_simulation_comparison}
\end{figure}

\begin{figure}[t]  
    \centering
    \begin{subfigure}[b]{0.32\textwidth}
        \centering
        \includegraphics[width=\textwidth]{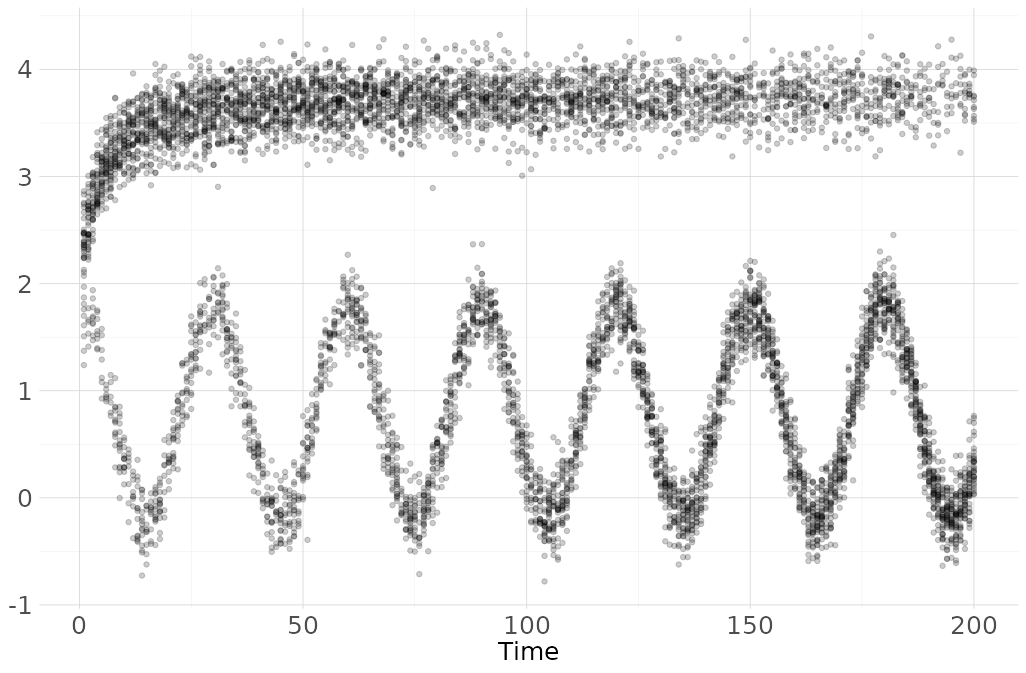}
        \caption{No overlap}
    \end{subfigure}
    \begin{subfigure}[b]{0.32\textwidth}
        \centering
        \includegraphics[width=\textwidth]{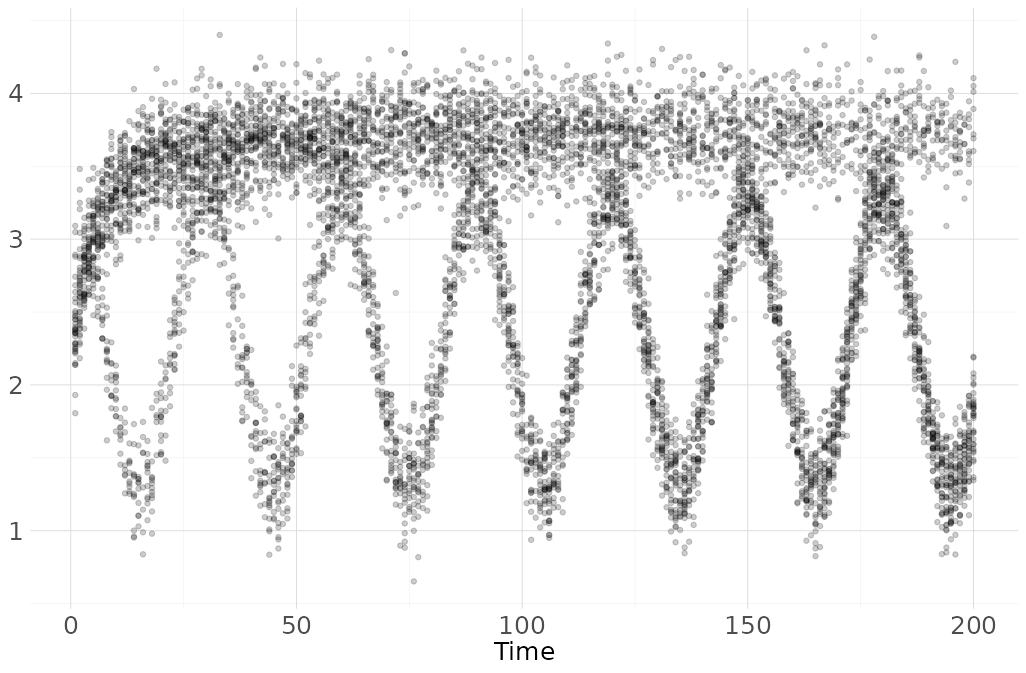}
        \caption{Mild overlap}
    \end{subfigure}
    \begin{subfigure}[b]{0.32\textwidth}
        \centering
        \includegraphics[width=\textwidth]{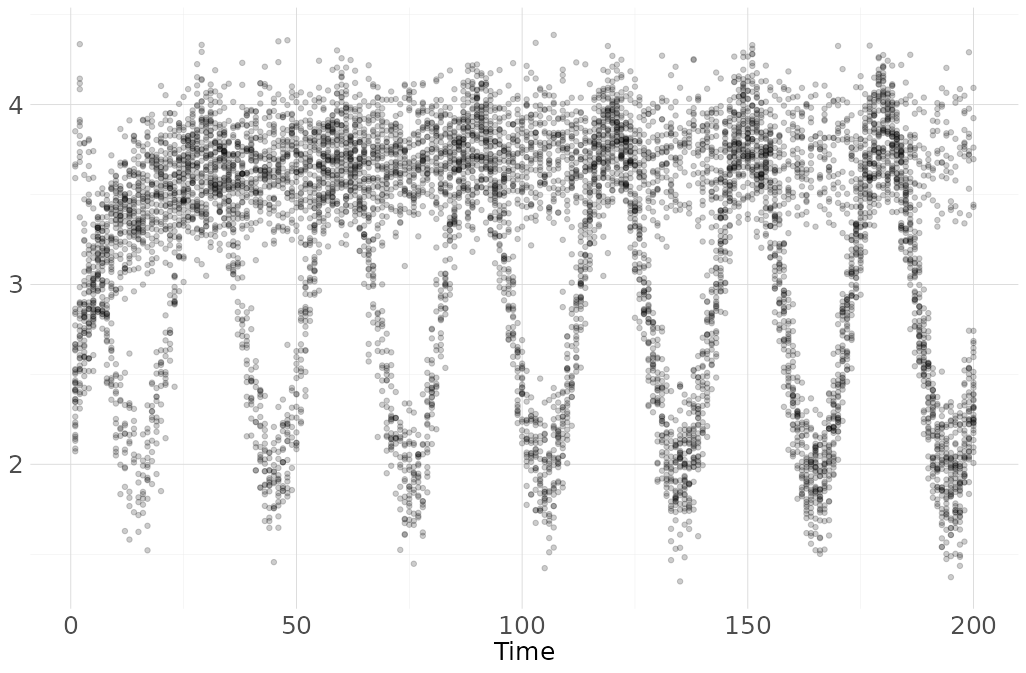}
        \caption{Considerable overlap}
    \end{subfigure}

    \caption{We simulate a gradually increasing amount of overlap between the clusters.}
    \label{fig:intersection-raw}
\end{figure}

We experiment first with cluster disappearance in the one-dimensional case, with a disappearance duration that gradually increases, as shown in Figure~\ref{fig:dis-duration-raw}. To get an idea of what exactly each method is doing, we examine the different fits at disappearance duration 20 in Figure~\ref{fig:disappearance_simulation_comparison}. We note that even though the missing cluster's mean drifts toward the other cluster in the kernel-EM algorithm in Figure~\ref{fig:kernel_fit_dis_20}, the cluster assignments did not change much because $\pi_{tk}$ is also smoothed out. The missing cluster's mean abruptly jumps to the other cluster's data immediately in the Hungarian fit in Figure~\ref{fig:hung_fit_dis_20}, leading to incorrect cluster assignments.  We also note that while the constant fit seems to do reasonably well with respect to cluster assignments in Figure~\ref{fig:const_fit_dis_20}, it is unable to capture the sinusoidal movement of the second cluster's mean at all. Figure~\ref{fig:intersection-raw} shows us the second simulation scenario, where clusters come closer together and eventually overlap.

\begin{figure}[t]
    \centering
    \begin{subfigure}[t]{0.48\textwidth}
        \centering
        \includegraphics[width=\linewidth]{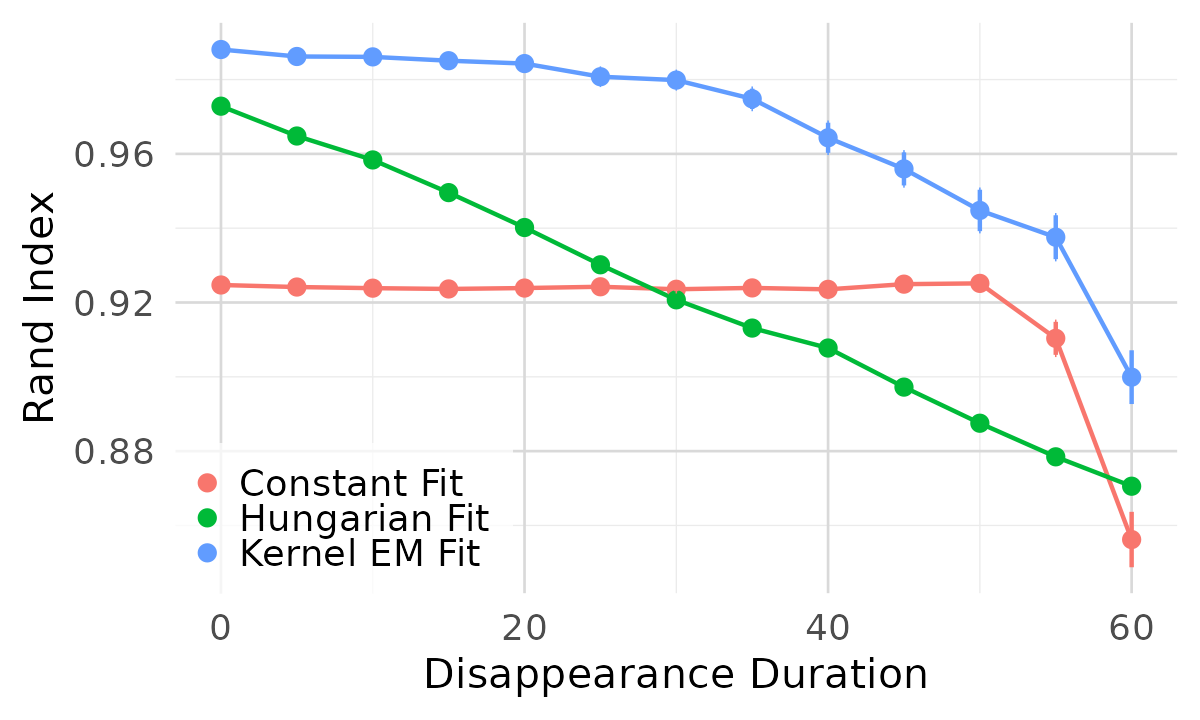}
        \caption{Disappearance simulation}
        \label{fig:rand_index_disappear}
    \end{subfigure}
    \hfill
    \begin{subfigure}[t]{0.48\textwidth}
        \centering
        \includegraphics[width=\linewidth]{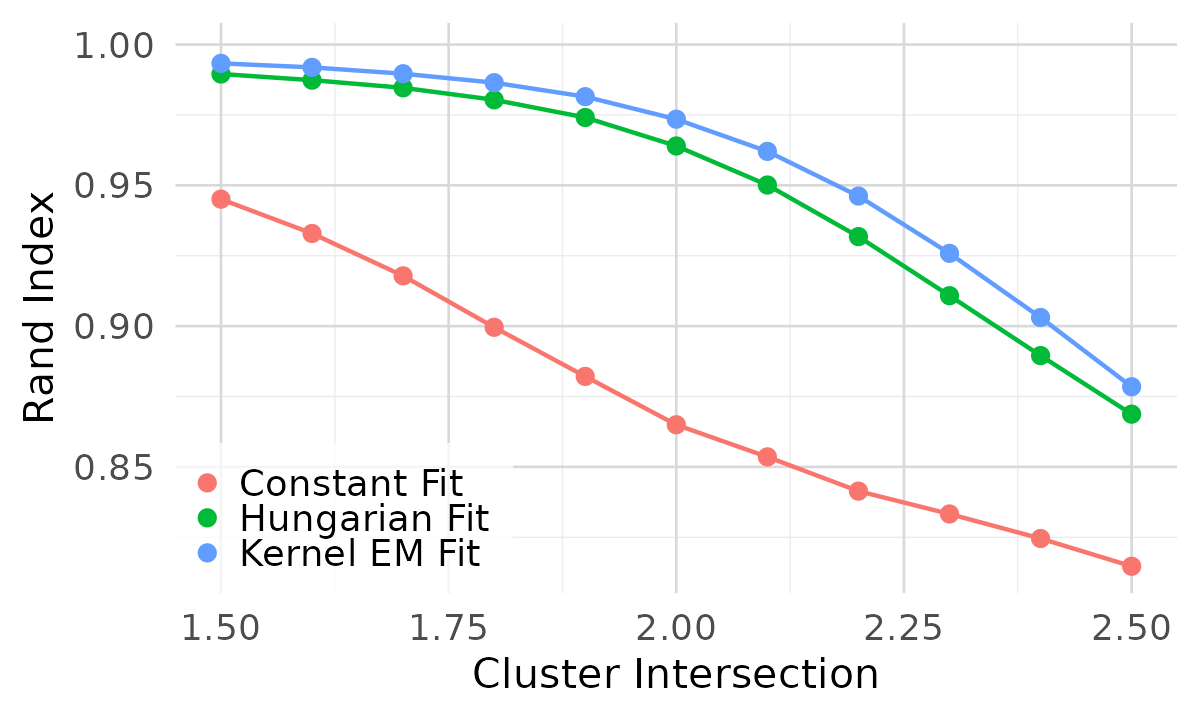}
        \caption{Intersect simulation}
        \label{fig:rand_index_intersect}
    \end{subfigure}
    \caption{Rand index comparison of constant fit, Hungarian matching, and kernel-EM fit across two simulation setups.}
    \label{fig:rand_index_comparison}
\end{figure}

Figures~\ref{fig:rand_index_disappear} and \ref{fig:rand_index_intersect} show the Rand index for all three methods under the two scenarios, where 100 simulations were run for every point plotted. We see that the problem becomes more challenging for all methods as either duration of disappearance or cluster overlap increases.  The kernel-EM method performs better than the other two methods across all disappearance durations and all levels of cluster intersection. 

%% file: 5-RealData.tex
\section{Real Data}
\label{sec:realdata}

We now evaluate the performance of our algorithm on real-world flow cytometry data from the SeaFlow dataset (\url{https://doi.org/10.5281/zenodo.2678021}). Over the last 15 years, SeaFlow has collected extensive high-resolution microbial data across 250,000 km, representing 800 billion cells collected from about 100 cruises \citep{ribalet2019seaflow}. SeaFlow records three optical properties: forward light scatter (fsc), chlorophyll fluorescence (692/40-nm bandpass filter), and phycoerythrin fluorescence (572/28-nm bandpass filter)\citep{swalwell2011seaflow}.  Additionally, forward light scatter is converted to cell size, represented as equivalent spherical diameter, and carbon biomass of each phytoplankton cell \citep{ribalet2019seaflow}. 

In this section, we examine a 2-week cruise in the spring of 2017 (MGL1704 cruise) that crossed the boundaries of the North Pacific Subtropical gyre where the composition of phytoplankton community varies dramatically \citep{dutkiewicz_multiple_2024, winter_shifts_2025}. Readings were acquired at approximately three-minute intervals over a two-week span (May 28 to June 13, 2017), with intermittent breaks, yielding a total of 355 hours of observation. On average, about 150,000 particles were detected per hour, resulting in an aggregate of 54 million detections.  

\begin{figure}[t]
    \centering
    \includegraphics[width=0.9\textwidth]{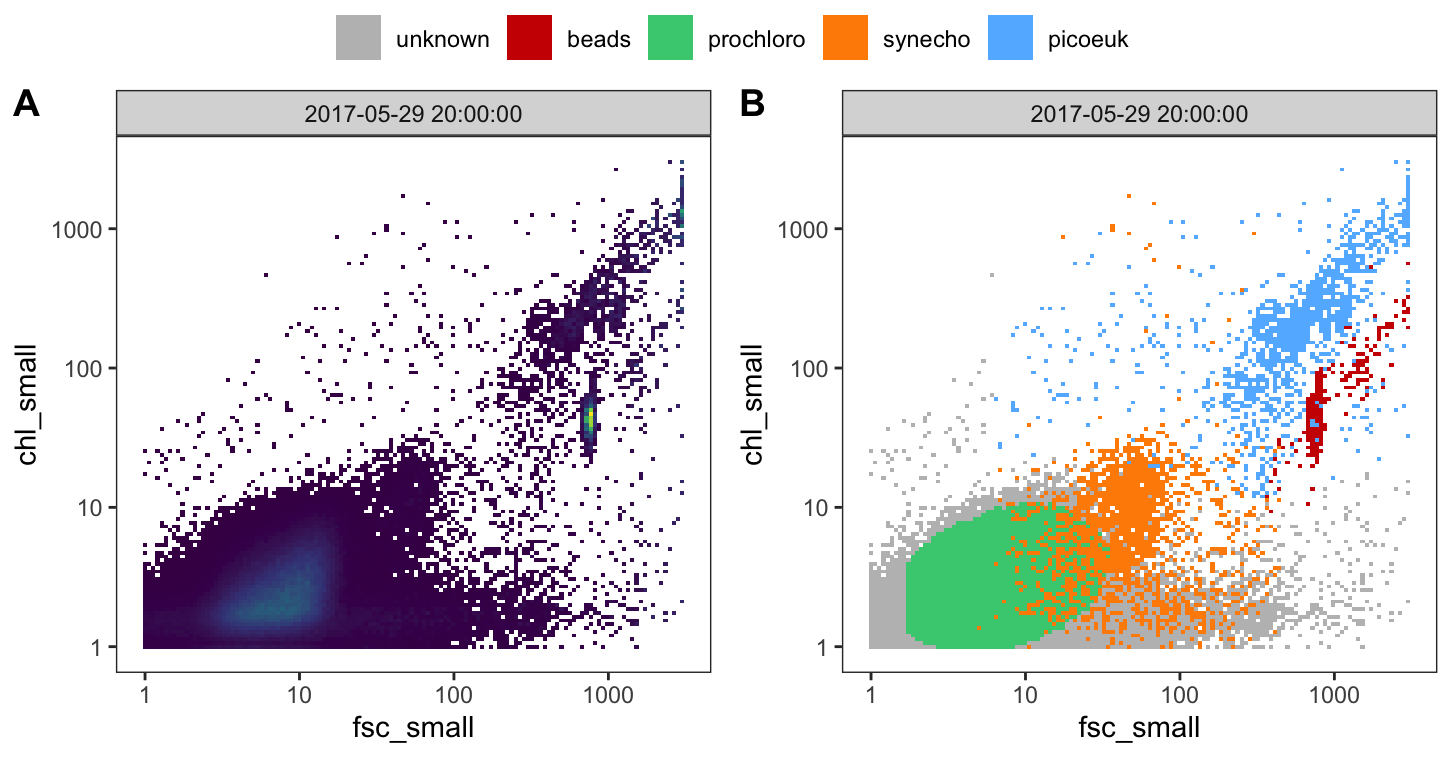}
    \caption{Manual gating of phytoplankton populations from flow cytometry data. Bivariate plots showing forward scatter (fsc\_small) versus chlorophyll fluorescence (chl\_small) from MGL1704 cruise data collected on May 29, 2017 at 20:00 (UTC). Raw data distribution prior to manual gating (A) and after manual population assignment (B), with distinct phytoplankton groups color-coded: \textit{Prochlorococcus} (green), \textit{Synechococcus} (orange), \textit{picoeukaryotes} (blue), reference beads (red), and unclassified particles (gray).}
    \label{fig:manual-gating-2-d}
\end{figure}

Raw SeaFlow data were manually gated using the combination of forward light scatter, chlorophyll fluorescence, and phycoerythrin fluorescence to identify four distinct populations: \textit{Prochlorococcus}, the smallest and most abundant phytoplankton in the ocean \citep{partensky_prochlorococcus_2010}, characterized by low chlorophyll fluorescence and detected near the instrument's background noise level; \textit{Synechococcus}, distinguished by their characteristic phycoerythrin fluorescence, giving them an orange fluorescence signature that separates them from other cyanobacteria; \textit{picoeukaryotes}, which are a diverse assemblage of small eukaryotic phytoplankton; and  1-$\mu$m yellow-green Fluoresbrite beads that are continuously injected into the sample stream as internal standards.  Particles that did not align with any of these populations were classified as “unknown”. Figure~\ref{fig:manual-gating-2-d} shows a two-dimensional slice of the cytometry data based on forward scatter and chlorophyll fluorescence only; phycoerythrin fluorescence was also used to gate, but it is not shown in these particular bivariate plots.

\begin{figure}[t]

    \centering
    \begin{subfigure}[t]{0.45\textwidth}
        \centering
        \includegraphics[width=\textwidth]{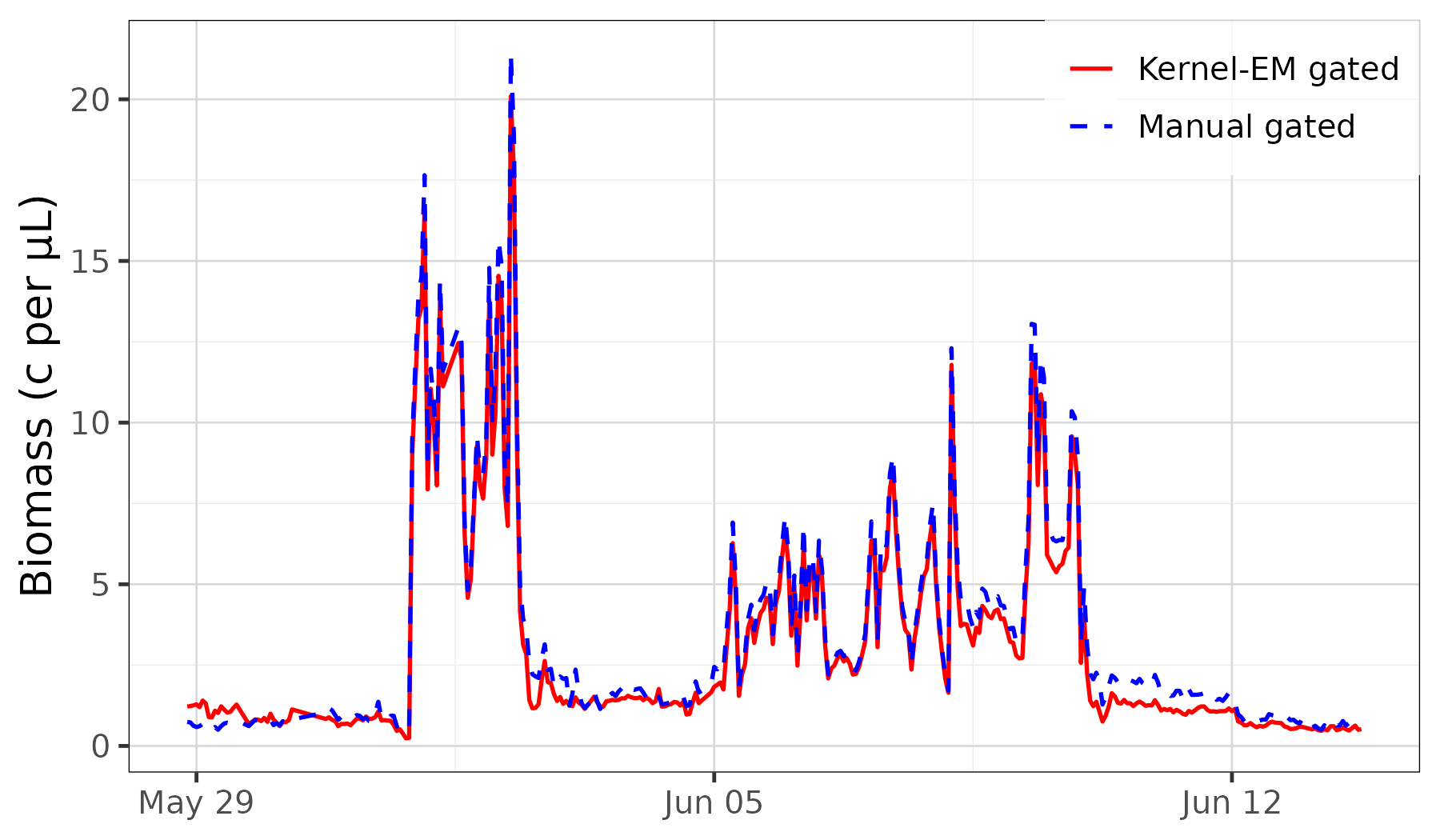}
        \caption{Synechococcus and Cluster 5}
    \end{subfigure}
    \hfill
    \begin{subfigure}[t]{0.45\textwidth}
        \centering
        \includegraphics[width=\textwidth]{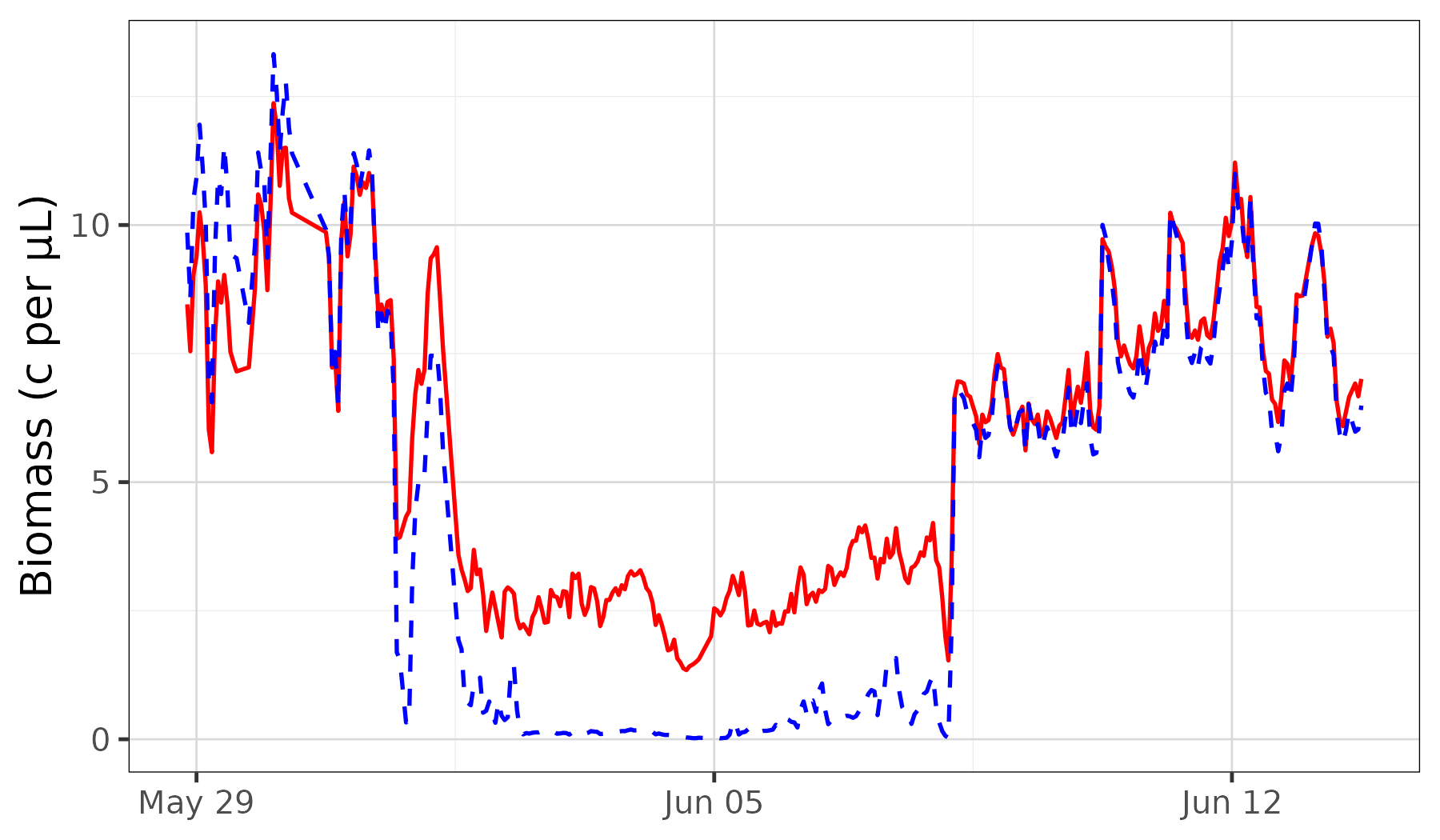}
        \caption{Prochlorococcus and Clusters 1 and 8}
    \end{subfigure}
    
    \vspace{0.3cm}
    
    \begin{subfigure}[t]{0.45\textwidth}
        \centering
        \includegraphics[width=\textwidth]{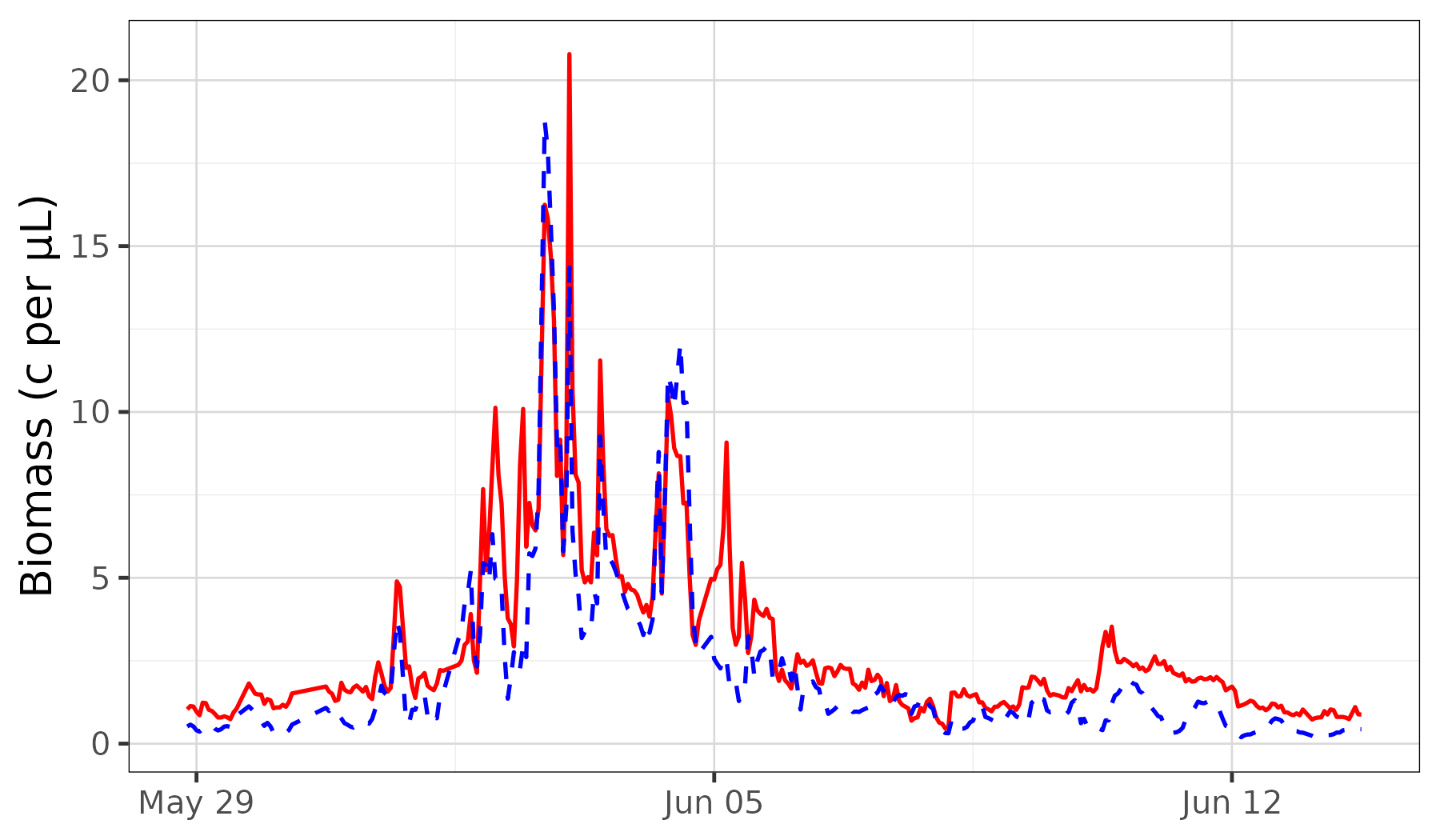}
        \caption{Picoeukaryotes and Cluster 6}
    \end{subfigure}
    \hfill
    \begin{subfigure}[t]{0.45\textwidth}
        \centering
        \includegraphics[width=\textwidth]{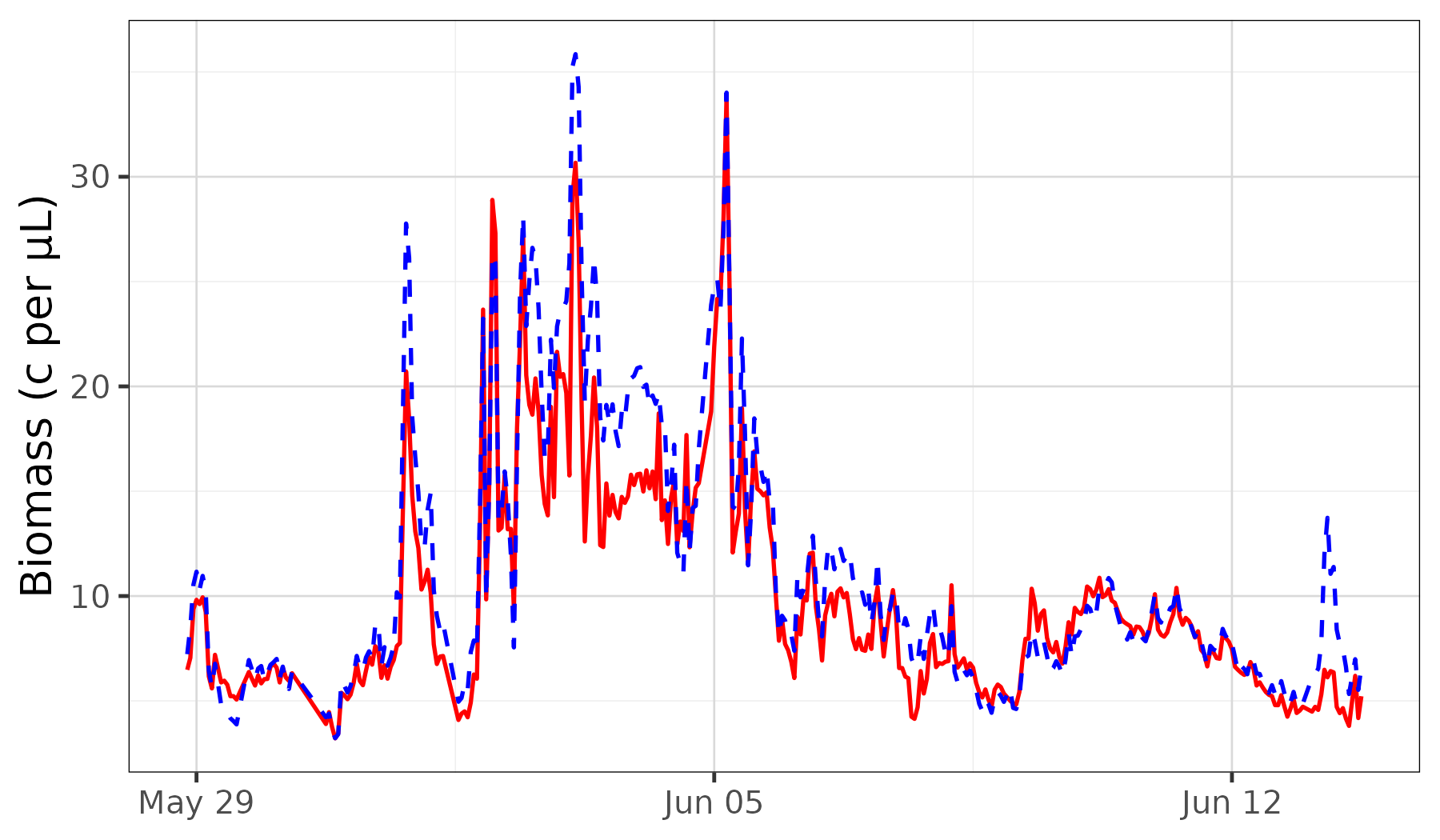}
        \caption{All Other Populations and Clusters 2, 3, 4, and 7}
    \end{subfigure}
    
    \caption{Total biomass over time for various phytoplankton populations from the MGL1704 cruise. The model was fit with $K=8$.}
    \label{fig:cytogram_plots_grid}
\end{figure}

To compare between the kernel-EM algorithm and manual gating, we start by looking at how the total biomass for each phytoplankton subpopulation changes over time, according to the two gating methods. Figure~\ref{fig:cytogram_plots_grid} shows this comparison for the 2017 cruise, fit using $K = 8$. We see that combinations of the automatically gated clusters give us total biomasses that match very closely with the population biomasses from manual gating. While this agreement with manual gating  is encouraging, it is important to recognize that our approach allows us to overcome several fundamental limitations inherent in manual cytogram interpretation. 

Firstly, human operators are unable to accurately cluster in three dimensions, and are therefore constrained to analyzing two-dimensional projections---as seen in Figure~\ref{fig:manual-gating-2-d}---of the three-dimensional cytogram space. This limitation forces operators to examine multiple two-dimensional scatter plots (e.g., forward scatter vs. chlorophyll fluorescence, forward scatter vs. phycoerythrin fluorescence), potentially missing important population boundaries that are only apparent when all three optical properties are considered simultaneously. Our automated approach leverages the full three-dimensional information content, enabling more precise population delineation.

Secondly, manual inspection is practically limited to identifying 3-4 distinct clusters by eye, as visual discrimination becomes increasingly difficult with higher cluster numbers. This constraint can lead to the overlooking of smaller but ecologically significant populations or the inappropriate lumping of distinct subpopulations. In Figure~\ref{fig:cytogram_plots_grid}, this limitation is evident where clusters 2, 3, 4, and 7 from our 8-cluster model are necessarily combined into a single "All Other" category for comparison with manual gating results from the 2017 cruise.

Lastly, manual gating applied independently to each time point often lacks temporal coherence, leading to discontinuous population trajectories that may not reflect actual biological dynamics. This challenge is particularly pronounced for populations that appear and disappear over time, such as \textit{Prochlorococcus} during the middle of the 2017 cruise, where manual gating struggles to maintain consistent identification criteria as population abundance fluctuates near detection thresholds, as seen in Figure~\ref{fig:cytogram_plots_grid}b. Our approach ensures that population changes are smooth, producing more biologically plausible temporal patterns that can track populations even through periods of low abundance or optical ambiguity. 

Another way to compare manual and automated gating is to run the algorithm directly on particle level data that has been manually gated, and sum up each cluster's responsibilities for a given manually assigned population. If we normalize over each manually-identified phytoplankton population, we get the confusion matrix in Table~\ref{tab:MGL1704_confusion_matrix}, showing us what fraction of each population belongs to each algorithm-generated cluster.

\begin{table}[H]
\centering
\begin{tabular}[t]{l|H|H|H|H|H}
\toprule
  & \multicolumn{1}{c|}{Beads}
  & \multicolumn{1}{c|}{Picoeukaryotes}
  & \multicolumn{1}{c|}{Prochlorococcus}
  & \multicolumn{1}{c|}{Synechococcus}
  & \multicolumn{1}{c}{Unknown}\\
\midrule
Cluster 1 & 0.00 & 0.00 & 0.53 & 0.00 & 0.49\\
Cluster 2 & 0.00 & 0.00 & 0.00 & 0.03 & 0.04\\
Cluster 3 & 1.00 & 0.13 & 0.00 & 0.01 & 0.00\\
Cluster 4 & 0.00 & 0.00 & 0.44 & 0.00 & 0.13\\
Cluster 5 & 0.00 & 0.75 & 0.00 & 0.00 & 0.02\\
Cluster 6 & 0.00 & 0.00 & 0.01 & 0.89 & 0.02\\
Cluster 7 & 0.00 & 0.00 & 0.03 & 0.01 & 0.30\\
Cluster 8 & 0.00 & 0.11 & 0.00 & 0.07 & 0.00\\
\bottomrule
\end{tabular}
\caption{Confusion matrix for the 2017 cruise with $K=8$. We normalize over each manually-identified phytoplankton population (making the columns sum to 1), showing how each manually gated population is captured by the automated gating.}
\label{tab:MGL1704_confusion_matrix}
\end{table}

\begin{figure}[t]
    \centering
    \makebox[\textwidth]{%
      \includegraphics[height=0.26in]{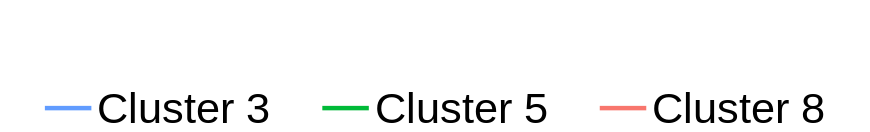}%
    }
    \vspace{-6pt} 

    \begin{subfigure}[t]{0.30\textwidth}
        \centering
        \includegraphics[width=\textwidth]{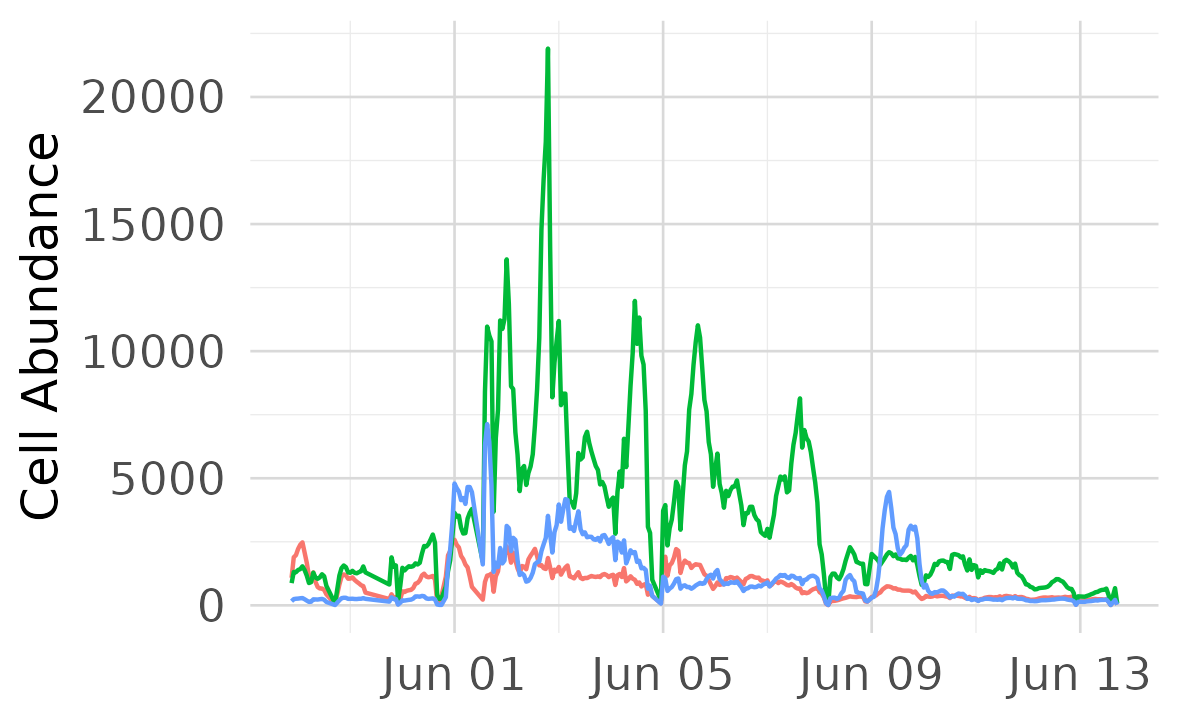}
        \caption{Cell Abundance}
    \end{subfigure}\hfill
    \begin{subfigure}[t]{0.30\textwidth}
        \centering
        \includegraphics[width=\textwidth]{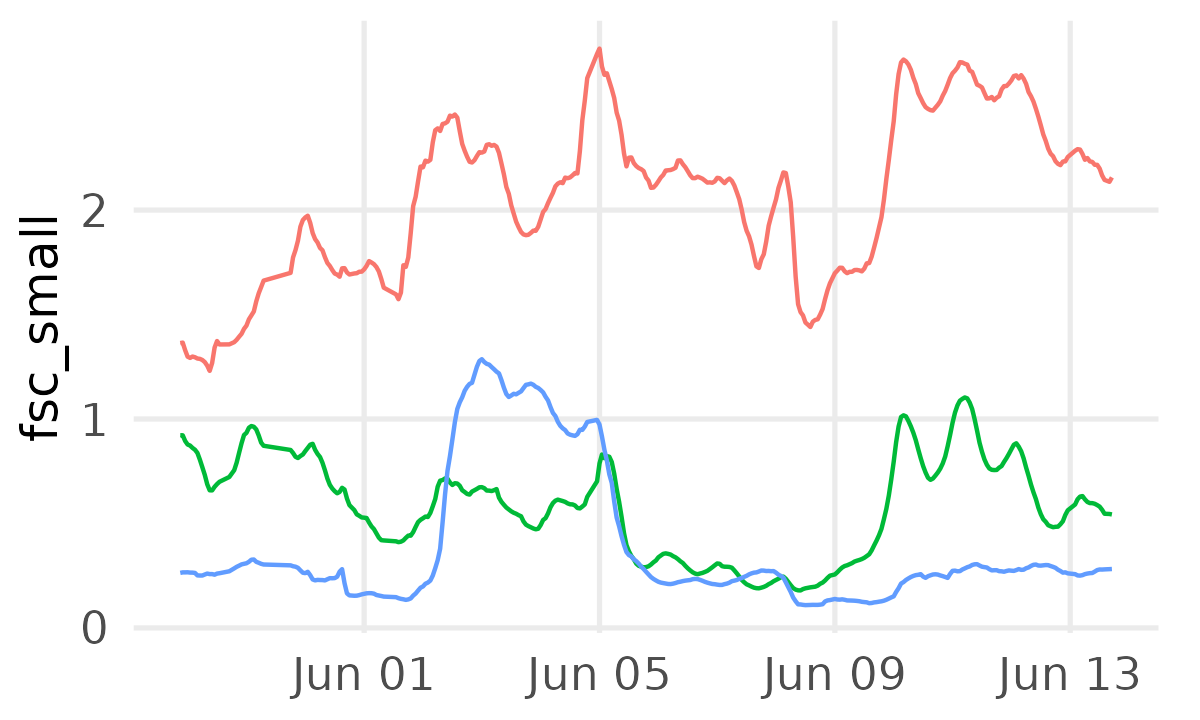}
        \caption{Forward Scatter}
    \end{subfigure}\hfill
    \begin{subfigure}[t]{0.30\textwidth}
        \centering
        \includegraphics[width=\textwidth]{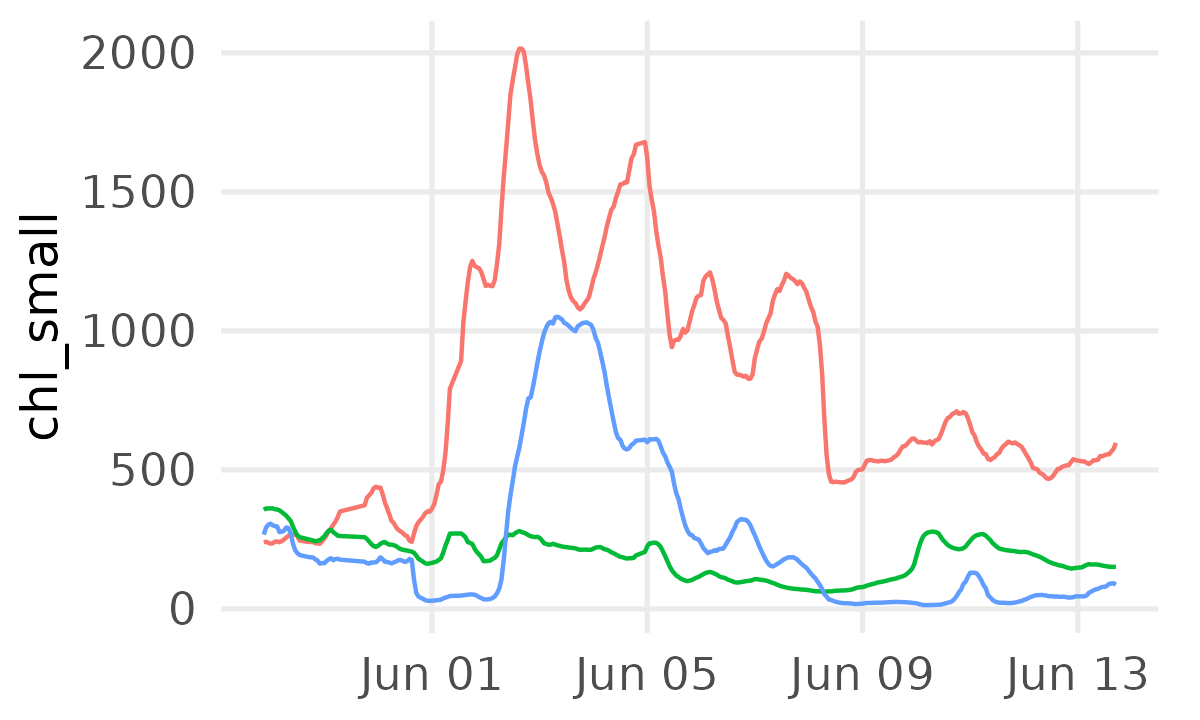}
        \caption{Chlorophyll‐Related Fluorescence}
    \end{subfigure}

    \caption{Picoeukaryote subpopulation summary: (a) cell abundance and (b–c) cluster centers over time.}
    \label{fig:pico-cluster-summary}
\end{figure}

The confusion matrix reveals how our automated approach partitions manually-gated populations into distinct subgroups: \textit{picoeukaryotes} distribute 75\% of their cells in cluster 5, 13\% in cluster 3, and 11\% in cluster 8, while all the reference beads are perfectly captured by cluster 3, and 89\% of the \textit{Synechococcus} population is captured by cluster 6. This clustering enables deeper analysis of subpopulation dynamics, as demonstrated in Figure~\ref{fig:pico-cluster-summary}, which tracks how the cell abundances, forward scatter (proxy of cell size), and chlorophyll content  of the \textit{picoeukaryote} subpopulations vary over time for the three picoeukaryote clusters. Cluster 5 (green) shows the highest abundance with a pronounced increase from June 1 to June 8, while cluster 8 (red) maintains consistently low abundance but contains the largest cells. Clusters 3 (blue) and 5 (green) contain similarly-sized cells, but their abundances peak at different times---cluster 3 peaks around June 3--4 while cluster 5 peaks toward the end of the cruise---indicating distinct responses to environmental gradients. 
These distinct temporal patterns suggest that what manual classification treats as a homogeneous population actually comprises multiple subgroups with different ecological characteristics. Each cluster likely represents distinct taxonomic groups with unique physiological responses to changing ocean conditions, demonstrating how automated clustering can reveal hidden structure in biological data and provide insights into how different organisms respond to environmental variation.

\captionsetup[subfigure]{aboveskip=2pt,belowskip=2pt}

%% file: 6-Discussions.tex
\section{Discussion}

In this work, we introduce a novel kernel-smoothed EM algorithm designed specifically to cluster data that evolves dynamically over time. Our approach has a number of advantages over manual gating: (i) the algorithm operates directly on the three-dimensional cytograms (whereas human curators generally look at two-dimensional projections), (ii) it takes as input many cytograms rather than making judgments using only one cytogram at a time, (iii) it is able to produce a higher number of clusters than a human would be able to discern, (iv) it offers a principled and reproducible framework that can be easily shared and scaled across multiple research groups. Through extensive simulation, we demonstrate that our proposed method outperforms other natural alternatives, providing more accurate and reliable clustering results. Additionally, when applied to real-world oceanographic data, the algorithm exhibits strong agreement with manually gated classifications while also revealing previously overlooked subpopulations, offering new insights into the structure of the data. 

Perhaps most significantly, the computational efficiency of our kernel-smoothed EM algorithm makes it particularly valuable for processing the massive datasets generated by modern flow cytometry systems, like SeaFlow. As autonomous platforms equipped with flow cytometers become more common, the ability to automatically process data from multiple instruments across ocean basins becomes essential for global-scale oceanographic research. Our method provides the foundation for such large-scale automated analysis systems.

Our approach also has potential applications beyond flow cytometry in marine systems: the fundamental principle of encouraging smooth parameter evolution over time applies broadly to temporal clustering problems.
In medical applications, for example, tracking immune cell populations over time during disease progression or treatment could benefit from the temporal coherence provided by kernel smoothing. Industrial applications might include monitoring product quality through time-evolving cluster analysis of manufacturing parameters.

There are still several future directions to pursue. From a theoretical perspective, it is well known that the EM algorithm has non-decreasing iterates of the likelihood at every iteration. Our kernel-EM algorithm was proposed as a modification of the EM algorithm and does not necessarily optimize an objective.  A key question is whether such an objective can be identified.  Besides adding to the theoretical understanding of our procedure, doing so could also be useful in determining stopping criteria and choosing solutions from a collection of multiple random initializations.

%% file: 001-acknowledgment.tex
This work was supported by grants from the Simons Foundation: the Simons Collaboration on Computational Biogeochemical Modeling of Marine Ecosystems/CBIOMES (Grant ID: 549939 to JB) and the Simons Microbial Oceanography Project Award (Grant ID: 574495 to FR). The authors acknowledge the Center for Advanced Research Computing (CARC) at the University of Southern California for providing computing resources that have contributed to the research results reported within this publication. URL: \url{https://carc.usc.edu}.

%% file: 7-Appendix.tex
\section{Appendix} \label{sec:appendix}

In this appendix, we have two sections: Section~\ref{subsec:bayesian_init} discusses a Bayesian approach to initialize our main kernel-smoothed EM Algorithm, and Section~\ref{subsec:thm1-proof} presents the proof of Theorem~\ref{thm:mse_comparison}. 

\subsection{Bayesian Initialization} \label{subsec:bayesian_init}

While the constant initialization described in Section~\ref{subsection:const-init} offers a quick start to our main algorithm, the EM-inspired Bayesian approach described in this section provides a more principled initialization that gives somewhat smooth estimates of our parameters over time. The process starts with running a standard EM algorithm for the first time point. For each subsequent time point $t$,  we perform a standard E-step, computing responsibilities $\gamma_{itk}$ based on the estimates $\hat\mu_{t-1, k},\hat\Sigma_{t-1, k}$, and $\hat\pi_{t-1, k}$ from the previous time point. We then proceed to estimate $\hat\mu_{tk}$, $\hat\Sigma_{tk}$ and $\hat\pi_{tk}$ using posterior means, assuming the appropriate conjugate priors for each parameter.

We let $n_{tk} = \sum_{i=1}^{n_{t}} \gamma_{itk}$ represent the total number of points in cluster $k$ at time $t$, where $n_t$ is  the total number of points across all clusters at time $t$. For the cluster means, for example, we assume a prior $\mu_{tk} \sim \mathcal{N} \left( \mu_{t-1, k}, \frac{ \Sigma_{t-1, k} }{ n_{t-1, k} } \right)$. Additionally, we treat $\hat\Sigma_{t-1, k}$ as the known covariance in the likelihood for time $t$, with $Y_{it} \mid \mu_{tk} \sim \mathcal{N}( \mu_{tk}, \Sigma_{t-1, k} ), \text{ for } i = 1, \dots, n_t$. 

We follow similar procedures for $\hat\Sigma_{tk}$ and $\hat\pi_{tk}$, selecting priors based on parameters at time point $t-1$, and updating the parameters at time $t$ using their posterior means. The full details of this procedure for all three parameters are provided in Section~\ref{subsubsection:bayes-pi-update}, Section~\ref{subsubsection:bayes-mu-update}, and Section~\ref{subsubsection:bayes-Sigma-update}, but we first examine the final update equations:

\begin{equation}\label{eq:Bayes-pi}
\hat{\pi}_{tk} = \frac{n_{t-1, k} + n_{t, k}}
{n_{t-1} + n_{t}},
\end{equation}

\begin{equation}\label{eq:Bayes-mu}
\hat{\mu}_{tk} = \frac{n_{t-1, k} \cdot \hat\mu_{kt-1} + \sum_{i=1}^{n_t} \gamma_{itk} Y_{it}}
{n_{t-1, k} + n_{t, k}},
\end{equation}

\begin{equation}\label{eq:Bayes-Sigma}
\hat{\Sigma}_{tk} = \frac{n_{t-1, k} \cdot \hat\Sigma_{kt-1} + \sum_{i=1}^{n_t} \gamma_{itk} \left(Y_{it} - \hat\mu_{tk}\right) \left(Y_{it} - \hat\mu_{tk}\right)^\top}
{n_{t-1, k} + n_{t, k}}.
\end{equation}

Note that the update for $\hat{\pi}_{tk}$ is obtained by summing the total number of points in cluster $k$ at times $t-1$ and $t$ and dividing by the combined total number of points across these two times. The updates for $\hat{\mu}_{tk}$ and $\hat{\Sigma}_{tk}$ are a weighted average between the parameters at the previous time $t-1$ and the MLE estimates at time $t$, based on the responsibilities $\gamma_{itk}$. The weights are determined by the total number of points assigned to each cluster at the respective time points. We now proceed to the detailed derivation of these three equations.

\subsubsection{\texorpdfstring{Posterior Mean for $\pi_{tk}$}{Posterior Mean for pi tk}}
\label{subsubsection:bayes-pi-update}

We model the cluster proportions $\boldsymbol{\pi}_t = (\pi_{t1}, \pi_{t2}, \dots, \pi_{tK})$ using a Dirichlet distribution. The effective prior counts for each cluster $k$ at time $t$ come from the responsibilities at time $t-1$: 
$$
\alpha_{tk} 
= 
\sum_{i=1}^{n_{t-1}} \gamma_{i,t-1,k}, 
$$
where $n_{t-1}$ is the total number of points at time $t-1$. We note that this notation entails $\alpha_{tk} = n_{t-1, k}$. Thus, the prior distribution of $\boldsymbol{\pi}_t$ is
$$ 
\boldsymbol{\pi}_t 
\sim 
\text{Dirichlet}(\boldsymbol{\alpha_t}), 
\quad 
\text{where} 
\quad 
\boldsymbol{\alpha_t} = (\alpha_{t1}, \alpha_{t2}, \dots, \alpha_{tK}).
$$
The effective counts for each cluster at time $t$ are $ n_{tk}  =  \sum_{i=1}^{n_t} \gamma_{i t k}$, which we assume follow a multinomial distribution:
$$
\mathbf{n_t} 
= 
(n_{t1}, n_{t2}, \dots, n_{tK}) 
\sim 
\text{Multinomial}\bigl(n_t, \boldsymbol{\pi}_t\bigr).
$$
 Using the conjugacy between the Dirichlet prior and the multinomial likelihood, the posterior distribution of $\boldsymbol{\pi}_t$ is
$$
\boldsymbol{\pi}_t 
\sim 
\text{Dirichlet}(\boldsymbol{\alpha_{t}} + \boldsymbol{n_{t}}).
$$
The posterior mean of $\pi_{tk}$ is then calculated as
$$
\hat{\pi}_{tk} 
= 
\frac{\alpha_{tk} + n_{tk}}{\sum_{k'=1}^K (\alpha_{tk'} + n_{tk'})}.
$$
Substituting the expressions for $\alpha_{tk}$ and $n_{tk}$ into the posterior mean, we get
$$
\hat{\pi}_{tk} 
= 
\frac{
  \displaystyle \sum_{i=1}^{n_{t-1}} \gamma_{i,t-1,k}
  \;+\;
  \sum_{i=1}^{n_t} \gamma_{i,t,k}
}{
  \displaystyle
  \sum_{k'=1}^K 
   \Bigl(
    \sum_{i=1}^{n_{t-1}} \gamma_{i,t-1,k'}
    \;+\;
    \sum_{i=1}^{n_t} \gamma_{i,t,k'}
   \Bigr)
}.
$$
Since the responsibilities sum to 1 over clusters for each data point, the denominator simplifies to $n_{t-1} + n_t$, resulting in the following final posterior mean updates:

\begin{equation} \label{eq:Bayes-pi-final}
\hat{\pi}_{tk} 
= 
\frac{
  \left(\sum_{i=1}^{n_{t-1}} \gamma_{i,t-1,k}\right) 
  \;+\; 
  \left(\sum_{i=1}^{n_t} \gamma_{i,t,k}\right)
}{
  n_{t-1} + n_t
} = \frac{
  n_{t-1, k} + n_{tk}
}{
  n_{t-1} + n_t
}.
\end{equation}

\subsubsection{\texorpdfstring{Posterior Mean for $\mu_{tk}$}{Posterior Mean for mu tk}}
\label{subsubsection:bayes-mu-update}

We now derive the posterior mean updates for the cluster means $\hat{\mu}_{tk}$. From time $t-1$ we have estimates $\hat\mu_{t-1, k}$ and $\hat\Sigma_{t-1, k}$, and we calculate the responsibilities $\hat\gamma_{i t k}$ at time $t$ based on these parameters from time $t-1$. We assume the prior distribution for $\mu_{tk}$ at time $t$ is

$$
\mu_{tk} \sim \mathcal{N} \left( \hat\mu_{t-1, k}, \frac{ \hat\Sigma_{t-1, k} }{ n_{t-1, k} } \right),
$$
where $
n_{t-1, k} = \sum_{i=1}^{n_{t-1}} \gamma_{i, t-1, k}
$ is the number of points in cluster $k$ at time $t-1$. We take the estimated covariance matrix $\hat\Sigma_{t-1, k}$ from the previous time point $t-1$ and use it as the covariance matrix in the likelihood function at time $t$:

$$
Y_{it} \mid \mu_{tk} \sim \mathcal{N}( \mu_{tk}, \hat\Sigma_{t-1, k} ), \quad i = 1, \dots, n_t.
$$
Using conjugacy, the posterior distribution of $\mu_{tk}$ is

$$
\mu_{tk} \mid Y_t \sim \mathcal{N} ( \hat{\mu}_{tk}, \hat\Sigma_{\text{post}}),
$$
where:

$$
\hat\Sigma_{\text{post}} = \left( \left( \frac{ \hat\Sigma_{t-1, k} }{ n_{t-1, k} } \right)^{-1} + n_{k t} \hat \Sigma_{t-1, k}^{-1} \right)^{-1} = \frac{ \hat\Sigma_{t-1, k} }{ n_{t-1, k} + n_{k t} },
$$

$$
\hat{\mu}_{tk} = \hat\Sigma_{\text{post}} \left( \left( \frac{ \hat \Sigma_{t-1, k} }{ n_{t-1, k} } \right)^{-1} \hat \mu_{t-1, k} + n_{k t} \hat\Sigma_{t-1, k}^{-1} \bar{Y}_{tk} \right ) = \frac{ n_{t-1, k} \hat \mu_{t-1, k} + n_{k t} \bar{Y}_{tk} }{ n_{t-1, k} + n_{k t} }.
$$
Here, $\bar{Y}_{tk} = \frac{\sum_{i=1}^{n_t} \gamma_{i t k} Y_{it} }{n_{k t}}$ is the cluster $k$ responsibility-weighted average of all points at time $t$. The posterior mean for $\hat{\mu}_{tk}$ expressed in terms of responsibilities gives us our final update:

\begin{equation} \label{eq:Bayes-mu-final}
\hat{\mu}_{tk} = \frac{ \left( \sum_{i=1}^{n_{t-1}} \gamma_{i, t-1, k} \right) \cdot \hat \mu_{t-1, k} + \sum_{i=1}^{n_t} \gamma_{i t k} Y_{it} }{ \left( \sum_{i=1}^{n_{t-1}} \gamma_{i, t-1, k} \right) + \left( \sum_{i=1}^{n_t} \gamma_{i t k} \right) } = \frac{n_{t-1, k} \cdot \hat\mu_{t-1, k} + \sum_{i=1}^{n_t} \gamma_{itk} Y_{it}}
{n_{t-1, k} + n_{t, k}}.
\end{equation}

\subsubsection{\texorpdfstring{Posterior Mean for $\Sigma_{tk}$}{Posterior Mean for Sigma tk}}
\label{subsubsection:bayes-Sigma-update}

We assume an inverse Wishart prior for $\Sigma_{tk}$:

$$
\Sigma_{tk} \sim \mathcal{IW} \left( \Psi_{t-1, k}, \nu_{t-1, k} \right),
$$
where

$$
\Psi_{t-1, k} =  n_{k , t -1}  \hat\Sigma_{t-1, k}, \quad \nu_{t-1, k} = n_{k , t -1} + d + 1,
$$
and $d$ is the dimension of the data. We choose $n_{k , t -1} + d + 1$ as the degrees of freedom to ensure the prior for $\Sigma_{t, k}$ is centered at $\hat\Sigma_{t-1, k}$:

$$
\mathbb{E}[\Sigma_{tk}] = \frac{ \Psi_{t-1, k} }{ \nu_{t-1, k} - d - 1 } = \frac{ n_{k , t -1}  \hat\Sigma_{t-1, k}}{ n_{k , t -1} + d + 1 - d - 1 } = \hat\Sigma_{t-1, k}.
$$
We then take $\hat\mu_{tk}$, which we estimated using the posterior mean as described above, and use it as the mean in the likelihood function at time $t$:

$$
Y_{it} \mid \Sigma_{tk} \sim \mathcal{N}( \hat\mu_{tk}, \Sigma_{tk} ), \quad i = 1, \dots, n_t.
$$
Using conjugacy, we know that the posterior distribution is also inverse Wishart, with posterior parameters

$$
\Psi_{tk} = \Psi_{t-1, k} + \sum_{i=1}^{n_t} \gamma_{i t k} \left( Y_{it} - \hat{\mu}_{tk} \right) \left( Y_{it} - \hat{\mu}_{tk} \right)^{T},
$$

$$
\nu_{tk} = \nu_{t-1, k} + n_{k t} = n_{k , t -1} + n_{k t} + d + 1.
$$
The posterior mean of $\Sigma_{tk}$ simplifies to give us the final update:

$$
\hat{\Sigma}_{tk} = \frac{ \Psi_{tk} }{ \nu_{tk} - d - 1 } = \frac{ n_{k , t -1}  \hat\Sigma_{t-1, k} + \sum_{i=1}^{n_t} \gamma_{i t k} \left( Y_{it} - \hat{\mu}_{tk} \right) \left( Y_{it} - \hat{\mu}_{tk} \right)^{T} }{n_{k , t -1} + n_{k t}}.
$$

\subsection{Proof of Theorem 1}\label{subsec:thm1-proof}

\subsubsection{MSE Result}

We will show the oracle estimator in the linear means scenario is unbiased, but the one in the absolute value scenario is biased, and since both estimators have the same variance, we will arrive at the MSE result.

We first recall that we assume $n = n_{t,0} = \sum_{i=1}^{n_t} \mathbf{1}\{ z_{it} = 0 \} = \sum_{i=1}^{n_t} \mathbf{1}\{ z_{it} = 1 \} = n_{t,1} $ for all $t$, which is the number of observations in each cluster at each time.

In the \textbf{Linear Means Scenario}, the cluster means are

$$
\mu_0^{lin}(t) = t, \quad \mu_1^{lin}(t) = -t.
$$
The oracle estimator for cluster 0 at time $ t $ is

$$
\hat{\mu}_0(t; Y, z^{lin}) = \frac{\displaystyle \sum_{s} w_{h}(t - s) \sum_{i=1}^{n_s} \mathbf{1}\{ z_{is}^{lin} = 0 \} Y_{is}}{\displaystyle \sum_{s} w_{h}(t - s) \sum_{i=1}^{n_s} \mathbf{1}\{ z_{is}^{lin} = 0 \}}.
$$
Using the fact that $ \mathbb{E}[Y_{is} \mid z_{is}^{lin} = 0] = \mu_0^{lin}(s) = s $, the expected value of $ \hat{\mu}_0(t; Y, z^{lin}) $ is

\begin{align*}
\mathbb{E}[\hat{\mu}_0(t; Y, z^{lin})] &= \frac{\displaystyle \sum_{s} w_{h}(t - s) \sum_{i=1}^{n_s} \mathbf{1}\{ z_{is}^{lin} = 0 \} \mathbb{E}[Y_{is} \mid z_{is}^{lin} = 0]}{\displaystyle \sum_{s} w_{h}(t - s) \sum_{i=1}^{n_s} \mathbf{1}\{ z_{is}^{lin} = 0 \}} \\
&= \frac{\displaystyle \sum_{s} w_{h}(t - s) n   s}{\displaystyle \sum_{s} w_{h}(t - s) n}.
\end{align*}
We define the weights

$$
\lambda_{t,s} = \frac{w_{h}(t - s)}{\displaystyle \sum_{s'} w_{h}(t - s') },
$$
which sum to 1 over $s$. Then

$$
\mathbb{E}[\hat{\mu}_0(t; Y, z^{lin})] = \sum_{s} \lambda_{t,s} s.
$$
Since the kernel weights $ w_{h}(t - s) $ are symmetric around $ t $ and $n$ is constant for all $s$, the weights $ \lambda_{t,s} $ are symmetric around $ t $. Therefore, the weighted average simplifies to

$$
\mathbb{E}[\hat{\mu}_0(t; Y, z^{lin})] = t = \mu_0^{lin}(t),
$$
showing that the estimator is \textbf{unbiased}. Similarly, it can be seen that $ \hat{\mu}_1(t; Y, z^{lin}) $ is unbiased.

In the \textbf{Absolute Value Scenario}, cluster means are given by

$$
\mu_0^{AV}(t) = |t|, \quad \mu_1^{AV}(t) = -|t|.
$$
The oracle estimator for cluster 0 at time $ t $ is

$$
\hat{\mu}_0(t; Y, z^{AV}) = \frac{\displaystyle \sum_{s} w_{h}(t - s) \sum_{i=1}^{n_s} \mathbf{1}\{ z_{is}^{AV} = 0 \} Y_{is}}{\displaystyle \sum_{s} w_{h}(t - s) \sum_{i=1}^{n_s} \mathbf{1}\{ z_{is}^{AV} = 0 \}}.
$$
Using $ \mathbb{E}[Y_{is} \mid z_{is}^{AV} = 0] = \mu_0^{AV}(s) = |s| $ and the same weights $ \lambda_{t,s} $ as before, the expected value of $ \hat{\mu}_0(t; Y, z^{AV}) $ is

$$
\mathbb{E}[\hat{\mu}_0(t; Y, z^{AV})] = \sum_{s} \lambda_{t,s} |s|.
$$
Due to the convexity of $ |s| $, we can apply Jensen's inequality:

$$
\mathbb{E}[\hat{\mu}_0(t; Y, z^{AV})] = \sum_{s} \lambda_{t,s} |s| \geq \left| \sum_{s} \lambda_{t,s} s \right| = |t| = \mu_0^{AV}(t).
$$
This shows that the estimator is \textbf{biased upward}.

Similarly, for cluster 1, we have

$$
\hat{\mu}_1(t; Y, z^{AV}) = \frac{\displaystyle \sum_{s} w_{h}(t - s) \sum_{i=1}^{n_s} \mathbf{1}\{ z_{is}^{AV} = 1 \} Y_{is}}{\displaystyle \sum_{s} w_{h}(t - s) \sum_{i=1}^{n_s} \mathbf{1}\{ z_{is}^{AV} = 1 \}},
$$
and using $ \mathbb{E}[Y_{is} \mid z_{is}^{AV} = 1] = \mu_1^{AV}(s) = -|s| $, we have

$$
\mathbb{E}[\hat{\mu}_1(t; Y, z^{AV})] = \sum_{s} \lambda_{t,s} (-|s|).
$$
Since $ -|s| $ is a concave function, applying Jensen's inequality gives us

$$
\mathbb{E}[\hat{\mu}_1(t; Y, z^{AV})] = \sum_{s} \lambda_{t,s} (-|s|) \leq -\left| \sum_{s} \lambda_{t,s} s \right| = -|t| = \mu_1^{AV}(t),
$$
showing us that the estimator is \textbf{biased downward}.

To show both estimators have the same variance, we begin by letting $\overline Y_{s,k}:=\frac{1}{n}\sum_{i=1}^n \mathbf{1}\{z_{is}=k\}\,Y_{is}$ be the within-$(s,k)$ sample mean. 
As $n_{s,k}\equiv n$ for all $s$ and $k$, the oracle estimator, for either labeling $z\in\{z^{lin},z^{AV}\}$, can be written as
\[
\hat\mu_k(t;Y,z)=\sum_{s}\alpha_{t,s}\,\overline Y_{s,k},\qquad 
\alpha_{t,s}:=\frac{w_h(t-s)}{\sum_{u} w_h(t-u)}.
\]
Since the noise variables $\varepsilon_{is}=Y_{is}-\mu_{Z_{is}}(s)$ are independent across $(i,s)$ with $\Var(\varepsilon_{is})=\sigma^2$, we have $\Var(Y_{is})=\sigma^2$ and hence $\Var(\overline Y_{s,k})=\sigma^2/n$ and $\Cov(\overline Y_{s,k},\overline Y_{u,k})=0$ for $s\neq u$.
Therefore,
\[
\Var\!\big(\hat\mu_k(t;Y,z)\big)
=\sum_{s}\alpha_{t,s}^2\,\Var(\overline Y_{s,k})
=\frac{\sigma^2}{n}\sum_{s}\alpha_{t,s}^2
=\frac{\sigma^2}{n}\,\frac{\sum_{s} w_h(t-s)^2}{\bigl(\sum_{u} w_h(t-u)\bigr)^2}.
\]

This depends only on the kernel weights and $n$—not on the mean functions $\mu^{lin}$ or \ $\mu^{AV}$ nor on the labeling $z^{lin}$ vs.\ $z^{AV}$—so both oracles have identical variance.

In summary, the estimator $ \hat{\mu}_k(t; Y, z^{AV}) $ is biased due to the convexity or concavity of the absolute value cluster means. Consequently, the MSE in absolute value scenario includes an additional bias term, leading to a higher MSE compared to the linear means scenario.

\subsubsection{Likelihood of a New Observation}

Let $\tilde{Y}_t$ be a new test point at time $t$. We have two equally valid models:

\begin{align}
\tilde{Y}_t &= \tilde{Z}_t^{lin} \mu_1^{lin}(t) + (1 - \tilde{Z}_t^{lin}) \mu_0^{lin}(t) + \tilde{\varepsilon}_t, \\
\tilde{Y}_t &= \tilde{Z}_t^{AV} \mu_1^{AV}(t) + (1 - \tilde{Z}_t^{AV}) \mu_0^{AV}(t) + \tilde{\varepsilon}_t,
\end{align}
where $\mathbb{P}(\tilde{Z}_t^{lin} = 0) = \mathbb{P}(\tilde{Z}_t^{lin} = 1) = \tfrac{1}{2}$, and similarly for $\tilde{Z}_t^{AV}$, and $\tilde{\varepsilon}_t \sim N(0, \sigma^2)$.

As the mixing proportions are equal and $\sigma^2$ is the same for both clusters, evaluating the likelihood of seeing the new point boils down to the distance of the new point from the cluster means. We therefore consider the expected sum of squared distances from each cluster, and will compute this quantity for the estimator from the \textbf{Linear Scenario} first:

\begin{align*}
\E\left[ \sum_k\left( \tilde{Y}_t - \hat{\mu}_{k}(t; Y, Z^{lin}) \right)^2 \right] = C_0^{lin} + C_1^{lin}.
\end{align*}
We will first compute $C_1^{lin}$, the term coming from cluster 1:

\begin{align*}
C_1^{lin} &= \E\left[ \left( \tilde{Y}_t - \hat{\mu}_{1}(t; Y, Z^{lin}) \right)^2 \right] \\&=\sum_{k=0}^{1} \mathbb{P}\left( \tilde{Z}_t^{lin} = k \right) \E\left[ \left( \tilde{Y}_t - \hat{\mu}_1(t; Y, Z^{lin}) \right)^2 \,\big|\, \tilde{Z}_t^{lin} = k \right] \\
&= \frac{1}{2} \sum_{k=0}^{1} \E\left[ \left( \mu_k^{lin}(t) + \tilde{\varepsilon}_t - \hat{\mu}_1(t; Y, Z^{lin}) \right)^2 \right]. \\
\end{align*}
Let's deal with the $k = 1$ term in this sum first: 
\begin{align*}
    \E\left[ \left( \mu_1^{lin}(t) + \tilde{\varepsilon}_t - \hat{\mu}_1(t; Y, Z^{lin}) \right)^2 \right] \
    &= \E\left[ \left( \mu_1^{lin}(t) - \hat{\mu}_1(t; Y, Z^{lin})\right)^2 + \tilde{\varepsilon}_t^2 + 2\tilde{\varepsilon}_t\left( \mu_1^{lin}(t) - \hat{\mu}_1(t; Y, Z^{lin})\right)    \right]\\ 
    &= \left[ \MSE(\hat{\mu}_1(t; Y, Z^{lin})) + \E\tilde{\varepsilon}_t^2 + 2\cdot \E\tilde{\varepsilon}_t \cdot\E\left( \mu_1^{lin}(t) - \hat{\mu}_1(t; Y, Z^{lin})\right)    \right]\\
    &= \MSE(\hat{\mu}_1(t; Y, Z^{lin})) +\sigma^2.
\end{align*}
We used the facts that $\tilde{\varepsilon}_t$ is independent of $\hat{\mu}_k(t; Y, Z^{lin})$, $\E\tilde{\varepsilon}_t^2 = \sigma^2$, and $\E\tilde{\varepsilon}_t = 0$. The next "cross cluster" term in the sum becomes

\begin{align*}
    \E\left[ \left( \mu_0^{lin}(t) + \tilde{\varepsilon}_t - \hat{\mu}_1(t; Y, Z^{lin}) \right)^2 \right] 
    &= \E\left[ \left( \mu_0^{lin}(t) - \mu_1^{lin}(t) + \mu_1^{lin}(t)- \hat{\mu}_1(t; Y, Z^{lin}) + \tilde{\varepsilon}_t  \right)^2 \right]\\
    &= (\mu_0^{lin}(t) - \mu_1^{lin}(t))^2 + \E \left[ \mu_1^{lin}(t) - \hat{\mu}_1(t; Y, Z^{lin})\right]^2 + \sigma^2 \\
    &= (2t)^2 + \MSE(\hat{\mu}_1(t; Y, Z^{lin}))+ \sigma^2.
\end{align*}
All the cross terms in the second step are zero because $\E\tilde{\varepsilon}_t = 0$ and $E \left[ \mu_1^{lin}(t) - \hat{\mu}_1(t; Y, Z^{lin})\right] = 0$ (the unbiased-ness shown above). We therefore get

\begin{align*}
C_1^{lin} &= \frac{1}{2}[\MSE(\hat{\mu}_1(t; Y, Z^{lin})) +\sigma^2 + (2t)^2 + \MSE(\hat{\mu}_1(t; Y, Z^{lin}))+ \sigma^2]\\
&=  \MSE(\hat{\mu}_1(t; Y, Z^{lin})) + \sigma^2 + 2t^2.
\end{align*}
Similarly, $C_0^{lin} = \MSE(\hat{\mu}_0(t; Y, Z^{lin})) + \sigma^2 + 2t^2$. Therefore, the expected sum of squared distances from each cluster in the linear scenario is

\begin{equation}
\label{eq:sq-dist-lin}
\E\left[ \sum_k\left( \tilde{Y}_t - \hat{\mu}_{k}(t; Y, Z^{lin}) \right)^2 \right] 
= \MSE(\hat{\mu}_1(t; Y, Z^{lin})) + \MSE(\hat{\mu}_0(t; Y, Z^{lin})) + 2\sigma^2 + 4t^2.
\end{equation}

In the \textbf{Absolute Value Scenario}, we have a similar situation to the one above, except that the estimators are biased and therefore have a higher MSE, and some cross terms that don't equal zero. We begin by breaking the expectation term-by-term as before:

\begin{align*}
\E\left[ \sum_k\left( \tilde{Y}_t - \hat{\mu}_{k}(t; Y, Z^{AV}) \right)^2 \right] = C_0^{AV} + C_1^{AV}.
\end{align*}
We will first compute $C_1^{AV}$, the term coming from cluster 1:

\begin{align*}
C_1^{AV} &= \E\left[ \left( \tilde{Y}_t - \hat{\mu}_{1}(t; Y, Z^{AV}) \right)^2 \right] \\&=\sum_{k=0}^{1} \mathbb{P}\left( \tilde{Z}_t^{AV} = k \right) \E\left[ \left( \tilde{Y}_t - \hat{\mu}_1(t; Y, Z^{AV}) \right)^2 \,\big|\, \tilde{Z}_t^{AV} = k \right] \\
&= \frac{1}{2} \sum_{k=0}^{1} \E\left[ \left( \mu_k^{AV}(t) + \tilde{\varepsilon}_t - \hat{\mu}_1(t; Y, Z^{AV}) \right)^2 \right]. \\
\end{align*}
The cluster-1 term simplifies as before: 
\begin{align*}
    \E\left[ \left( \mu_1^{AV}(t) + \tilde{\varepsilon}_t - \hat{\mu}_1(t; Y, Z^{AV}) \right)^2 \right] 
    &= \MSE(\hat{\mu}_1(t; Y, Z^{AV})) +\sigma^2.
\end{align*}
The next term, however, is where the two scenarios differ:

\begin{align*}
    \E\left[ \left( \mu_0^{AV}(t) + \tilde{\varepsilon}_t - \hat{\mu}_1(t; Y, Z^{AV}) \right)^2 \right] 
    &= \E\left[ \left( \mu_0^{AV}(t) - \mu_1^{AV}(t) + \mu_1^{AV}(t)- \hat{\mu}_1(t; Y, Z^{AV}) + \tilde{\varepsilon}_t  \right)^2 \right]\\
    &= (\mu_0^{AV}(t) - \mu_1^{AV}(t))^2 + \E \left[ \mu_1^{AV}(t) - \hat{\mu}_1(t; Y, Z^{AV})\right]^2 + \sigma^2 \\
    &\phantom{=} + 2 (\mu_0^{AV}(t) - \mu_1^{AV}(t))\E \left[ \mu_1^{AV}(t) - \hat{\mu}_1(t; Y, Z^{AV})\right]\\
    &= (2t)^2 + \MSE(\hat{\mu}_1(t; Y, Z^{AV}))+ \sigma^2 \\
    &\phantom{=} + 2 (\mu_0^{AV}(t) - \mu_1^{AV}(t))\E \left[ \mu_1^{AV}(t) - \hat{\mu}_1(t; Y, Z^{AV})\right]\\
    &\geq 4t^2 + \MSE(\hat{\mu}_1(t; Y, Z^{AV}))+ \sigma^2,
\end{align*}
where the last inequality follows from the fact that $(\mu_0^{AV}(t) - \mu_1^{AV}(t))$ is always positive, and $\E \left[ \mu_1^{AV}(t) - \hat{\mu}_1(t; Y, Z^{AV})\right]$ is also positive due to the downward bias of "trimming the hill", as shown earlier. We therefore get 

\begin{align*}
C_1^{AV} &\geq \frac{1}{2}[\MSE(\hat{\mu}_1(t; Y, Z^{AV})) +\sigma^2 + 4t^2 + \MSE(\hat{\mu}_1(t; Y, Z^{AV}))+ \sigma^2]\\
&=  \MSE(\hat{\mu}_1(t; Y, Z^{AV})) + \sigma^2 + 2t^2.
\end{align*}
Similarly,  $C^{AV}_0 \geq \MSE(\hat{\mu}_0(t; Y, Z^{AV})) + \sigma^2 + 2t^2 $. Therefore, for the expected sum of squared distances from each cluster in the absolute scenario, we have

\begin{equation}
\label{eq:sq-dist-av}
\E\left[ \sum_k\left( \tilde{Y}_t - \hat{\mu}_{k}(t; Y, Z^{AV}) \right)^2 \right] 
\geq \MSE(\hat{\mu}_1(t; Y, Z^{AV})) + \MSE(\hat{\mu}_0(t; Y, Z^{AV})) + 2\sigma^2 + 4t^2.
\end{equation}

Combining ~\eqref{eq:mse_inequality},~\eqref{eq:sq-dist-lin}, and ~\eqref{eq:sq-dist-av}, we finally get

\begin{align*}
\E\left[ \sum_k\left( \tilde{Y}_t - \hat{\mu}_{k}(t; Y, Z^{lin}) \right)^2 \right] &\stackrel{~\eqref{eq:sq-dist-lin}}{=}
 \MSE(\hat{\mu}_1(t; Y, Z^{lin})) + \MSE(\hat{\mu}_0(t; Y, Z^{lin})) + 2\sigma^2 + 4t^2 \\
&\stackrel{~\eqref{eq:mse_inequality}}{\leq} \MSE(\hat{\mu}_1(t; Y, Z^{AV})) + \MSE(\hat{\mu}_0(t; Y, Z^{AV})) + 2\sigma^2 + 4t^2 \\
&\stackrel{~\eqref{eq:sq-dist-av}}{\leq} \E\left[ \sum_k\left( \tilde{Y}_t - \hat{\mu}_{k}(t; Y, Z^{AV}) \right)^2 \right],
\end{align*}
which tells us that the estimator based on the linear cluster assignments predicts better on new data than the estimator based on the absolute value cluster assignments.